\documentclass[aps,english,preprint,nofootinbib]{revtex4}

\usepackage{graphicx}
\usepackage{hyperref}
\usepackage{amsmath, amssymb}
\usepackage{babel}
\usepackage{color}
\usepackage{mathrsfs}
\usepackage{slashed}

\begin{document}

\title{New method for calculating electromagnetic effects
	in semileptonic beta-decays of mesons }
%New lattice QCD strategies to calculate electromagnetic effects in semileptonic beta decays}

\author{Chien-Yeah Seng$^{1}$}
\author{Xu Feng$^{2,3,4,5}$}
\author{Mikhail Gorchtein$^{6,7,8}$}
\author{Lu-Chang Jin$^{9,10}$} 
\author{Ulf-G. Mei{\ss}ner$^{1,11,12}$}

\affiliation{$^{1}$Helmholtz-Institut f\"{u}r Strahlen- und Kernphysik and Bethe Center for Theoretical Physics,\\
	Universit\"{a}t Bonn, 53115 Bonn, Germany}
\affiliation{$^{2}$School of Physics, Peking University, Beijing 100871, China}
\affiliation{$^{3}$Collaborative Innovation Center of Quantum Matter, Beijing 100871, China}
\affiliation{$^{4}$Center for High Energy Physics, Peking University, Beijing 100871, China}
\affiliation{$^{5}$State Key Laboratory of Nuclear Physics and Technology, Peking University, Beijing 100871, China}
\affiliation{$^{6}$Helmholtz Institute Mainz, D-55099 Mainz, Germany}
\affiliation{$^{7}$GSI Helmholtzzentrum f\"ur Schwerionenforschung, 64291 Darmstadt, Germany}
\affiliation{$^{8}$Johannes Gutenberg University, D-55099 Mainz, Germany}
\affiliation{$^{9}$RIKEN-BNL Research Center, Brookhaven National Lab, Upton, NY, 11973, USA}
\affiliation{$^{10}$Physics Department, University of Connecticut, Storrs, Connecticut 06269-3046, USA}
\affiliation{$^{11}$Institute for Advanced Simulation, Institut f\"ur Kernphysik and J\"ulich Center for Hadron Physics, Forschungszentrum J\"ulich, 52425 J\"ulich, Germany}
\affiliation{$^{12}$Tbilisi State  University,  0186 Tbilisi, Georgia}

\date{\today}

\begin{abstract}

We construct several classes of hadronic matrix elements and relate them to the low-energy constants
in Chiral Perturbation Theory that describe the electromagnetic effects in the  semileptonic beta decay
of the pion and the kaon. We propose to calculate them using lattice QCD, and argue that such a
calculation will make an immediate impact to a number of interesting topics at the precision frontier,
including the outstanding anomalies in $|V_{us}|$ and the top-row Cabibbo-Kobayashi-Maskawa matrix
unitarity. 

\end{abstract}

\maketitle

%%%%%%%%%%%%%%%%%%%%%%%%%%%%%%%%%%%%%%%%%%%%%%%%%%%%%%%%%%%%%%%%%%%%
%\section{Introduction}
%%%%%%%%%%%%%%%%%%%%%%%%%%%%%%%%%%%%%%%%%%%%%%%%%%%%%%%%%%%%%%%%%%%%

\section{Introduction}

The last few years have seen a rapid development in the theory of the electroweak radiative corrections
(RCs) in hadron and nuclear beta decay processes. In particular, a dispersion relation
analysis~\cite{Seng:2018yzq,Seng:2018qru} significantly reduced the hadronic uncertainty of the
single-particle RCs in free neutron and superallowed nuclear beta decays, and led to a new status
of the top-row Cabibbo-Kobayashi-Maskawa (CKM) matrix unitarity, as quoted in the 2020 Particle
Data Group (PDG)~\cite{Zyla:2020zbs}:
\begin{equation}
|V_{ud}|^2+|V_{us}|^2+|V_{ub}|^2=0.9985(3)_{V_{ud}}(4)_{V_{us}},\label{eq:CKM2020}
\end{equation}
in contrast to the result in the 2018 PDG~\cite{Tanabashi:2018oca} with $0.9994(4)_{V_{ud}}(4)_{V_{us}}$
at the right hand side (RHS). The apparent violation of the top-row CKM unitarity at a 3$\sigma$
level and its implications on the possible physics Beyond the Standard Model (BSM)~\cite{Gonzalez-Alonso:2018omy,Bryman:2019ssi,Belfatto:2019swo,Cirgiliano:2019nyn,Tan:2019yqp,Bryman:2019bjg,Grossman:2019bzp,Coutinho:2019aiy,Cheung:2020vqm,Crivellin:2020lzu,Endo:2020tkb,Capdevila:2020rrl,Kirk:2020wdk} trigger renewed interest from both the experimental and theoretical community in the
precision frontier.

The improvements in the recent years mainly concern the reduction of the Standard Model (SM) theory
uncertainties in the extraction of $V_{ud}$. And now, as indicated in Eq.\eqref{eq:CKM2020}, the next
breakthrough must involve a similar reduction of the $V_{us}$ theory uncertainties. In particular,
the outstanding disagreement between the $V_{us}$ extracted from the kaon semileptonic decay
($K_{l3}$) and leptonic decay ($K_{l2}$)~\cite{Zyla:2020zbs}:
\begin{equation}
|V_{us}|=\left\{
\begin{array}{c}
0.2231(4)_\mathrm{exp+RCs}(6)_\mathrm{lattice}\:\:(N_f=2+1+1,\:K_{l3})\\
0.2252(5)\:\:(N_f=2+1+1,\:K_{\mu 2})
\end{array}\right.
\end{equation}
has to be understood. Apart from possible BSM explanations, such a
disagreement could originate either from unknown systematic errors in the SM input of the $K\pi$
form factor or, although somewhat less likely, the RCs in $K_{l3}$. For the first case one simply
needs a better lattice Quantum Chromodynamics (QCD) calculation of the $K\pi$ form factor
at zero momentum transfer, whereas the second case is much more complicated and will be the
focus in this paper. In particular, we will discuss the possible roles that lattice QCD can
play in this aspect.

Recently lattice QCD has made a tremendous progress in  first-principles studies of Quantum
Electrodynamics (QED) corrections to hadronic processes,
see e.g.~\cite{Carrasco:2015xwa,Lubicz:2016xro,Giusti:2017dwk,DiCarlo:2019thl}. In particular,
Ref.~\cite{Giusti:2017dwk} presented, for the first time, the full lattice study of the QED RCs to
the $K_{\mu 2}$ and $\pi_{\mu 2}$ decay rates, which involves a direct calculation of both the virtual
and real photon emission diagrams. The extension of the method above to semileptonic decay processes
is, however, expected to be extremely challenging~\cite{Giusti:2018guw,Sachrajda:2019uhh,Cirigliano:2019jig}. On the other hand, Ref.~\cite{Feng:2020zdc} adopted a completely different starting point, namely
to calculate the so-called ``axial $\gamma W$ box diagram'' on the lattice, which resulted in a
significant reduction of the theory uncertainty in $\pi_{e3}$~\cite{Feng:2020zdc}, and also
provided an independent cross-check of the dispersion relation analysis in the neutron
RCs~\cite{Seng:2020wjq}. This is the first time lattice QCD ever plays a decisive role in the
understanding of RCs of semi-leptonic beta decays, so a natural question to ask is whether the
same method is going to teach us anything useful about the RCs in $K_{l3}$, which is much
more complicated than $\pi_{e3}$ due to its larger Q-value.

The answer is yes if we appropriately combine lattice QCD with the existing theory framework. We
first recall that the standard approach to deal with the electroweak RCs in $K_{l3}$ is based on
Chiral Perturbation Theory (ChPT)~\cite{Cirigliano:2001mk,Cirigliano:2008wn}, in which the
theoretical uncertainties are from two sources: (1) the neglected terms that scale as higher-order
in the chiral power counting, and (2) the unknown low-energy constants (LECs). The first can
in principle be reduced by including higher-order loop corrections, whereas the second represents
a more fundamental issue: the LECs characterize the unknown dynamics of QCD at the chiral
symmetry breaking scale $\Lambda_\chi\sim 1$~GeV. The LECs are not constrained by chiral symmetry,
and there is no reliable experimental constraint on the ones that describe the electromagnetic
interactions of mesons. They are so far only calculated within
models~\cite{Ananthanarayan:2004qk,DescotesGenon:2005pw} with no rigorous error analysis.
Therefore, the ability to determine the relevant LECs with high accuracy will serve
as a first step in the breakthrough of the $V_{us}$ theory.

There is also another motivation to get more reliable values of these LECs. In leptonic decay
processes, one extracts  $|V_{us}/V_{ud}|$ by considering the ratio $R_A=\Gamma_{K_{\mu2}}/\Gamma_{\pi_{\mu2}}$~\cite{Marciano:2004uf}, because it turns out that the $K_{\mu 2}$ and $\pi_{\mu 2}$ decay rates share not only the
same short-distance electroweak RCs, but also the same combination of LECs at $\mathcal{O}(e^2p^2)$ so they
cancel out in the ratio. This leads to a smaller theoretical uncertainty than the extractions of the
individual $|V_{us}|$ and $|V_{ud}|$ themselves. Recently, a similar ratio $R_V=\Gamma_{K_{l3}}/\Gamma_{\pi_{e3}}$
was introduced for the semileptonic decay processes~\cite{Czarnecki:2019iwz}, which provides another venue
to extract $|V_{us}/V_{ud}|$ and could shed new lights on the $V_{us}$ discrepancy mentioned above. However,
we find that $\Gamma_{K_{l3}}$ and $\Gamma_{\pi_{e3}}$ do not share the same LECs at $\mathcal{O}(e^2p^2)$ and
so they do not fully cancel in the ratio. Therefore, one could better make use of $R_V$ if its residual dependence on the LECs can be fixed through an extra lattice QCD calculation.

In this paper we demonstrate how all the LECs relevant for the RCs in $K_{l3}$ and $\pi_{e3}$ can be pinned
down by calculating two types of rather simple hadronic matrix elements on lattice. The first type is
just the axial $\gamma W$ box diagram, which has already been done for pion. We derive a matching
relation between this quantity and the relevant LECs, and show that the lattice QCD result differs
significantly from the widely-adopted value based on resonance model estimation~\cite{DescotesGenon:2005pw},
which motivates us even further for a thorough re-analysis. A similar calculation of the $K_{e3}^0$ box
diagram at the SU(3) symmetric point will eventually fix all the needed LECs that describe the lepton-hadron
electromagnetic interactions. Finally, for the remaining LECs that do not involve a lepton, we
propose a lattice calculation of the four-point correlation functions based on the construction in
Ref.~\cite{Ananthanarayan:2004qk}. 

The contents in this paper are arranged as follows. In Sec.~\ref{sec:RC} we review the existing theory
frameworks to study the electroweak RCs in kaon and pion semileptonic decays, including the classical
``Sirlin's representation'' and the modern ChPT representation. We show in Sec.~\ref{sec:axialbox} that
comparing these two representations in the SU(3) limit
gives an elegant
matching relation between a subset of LECs and the axial $\gamma W$ box diagram calculable on lattice.
We discuss the implications of the lattice result in Ref.~\cite{Feng:2020zdc} and propose a similar
calculation in the $K\pi$ system. In Sec.~\ref{sec:4point} we construct a class of four-point correlation
functions that enable a direct lattice determination of the lepton-free LECs. Our final conclusions
are given in Sec.~\ref{sec:conclusion}.

\section{\label{sec:RC}Radiative corrections to semileptonic beta decays in two representations}

We start by reviewing the existing theoretical frameworks in the treatment of the semileptonic decay
of a generic spinless particle $\phi$, and its corresponding electroweak RCs. First,  the electromagnetic and
charged weak currents in the quark sector are defined as:
\begin{equation}
J_\mathrm{em}^\mu=\frac{2}{3}\bar{u}\gamma^\mu u-\frac{1}{3}\bar{d}\gamma^\mu d-\frac{1}{3}\bar{s}\gamma^\mu s,
\:\:\:J_W^\mu=V_{ud}\bar{u}\gamma^\mu(1-\gamma_5)d+V_{us}\bar{u}\gamma^\mu(1-\gamma_5)s,
\end{equation}
and the matrix element of the charged weak current can be expressed in terms of two form factors:
\begin{equation}
F^\mu_{fi}(p',p)=\left\langle \phi_f(p')\right|J_W^{\mu\dagger}(0)\left|\phi_i(p)\right\rangle=F_+^{fi}(t)
(p+p')^\mu+F_-^{fi}(t)(p-p')^\mu,
\end{equation}
where $t=(p-p')^2$. Notice that in the definition above the form factors contain the CKM matrix elements.
It is useful to remember that the contribution from $F_-^{fi}$ to the decay rate is suppressed at tree level
by the factor $m_l^2/M_{\phi_i}^2$, where $l$ is the emitted charged lepton. 

Now let us consider the decay process $\phi_i(p)\to \phi_f(p') e^+(p_e)\nu_e(p_\nu)$,
where $\phi_{i,f}$ are spinless particles. At tree level the decay amplitude is given by:
\begin{equation}
M_0=-\frac{G_F}{\sqrt{2}}\bar{u}_\nu\gamma_\lambda(1-\gamma_5)v_eF_{fi}^\lambda(p',p)~.
\end{equation}
Here, $G_F=1.1663787(6)\times 10^{-5}$~GeV$^{-2}$ is the Fermi constant measured in muon decay. This
definition has a natural advantage as it absorbs a large portion of the electroweak RCs that is common
to both the muon and hadron semileptonic beta decays into the definition of $G_F$.

Next we discuss the two different representations of the electroweak RCs in this decay process, namely
Sirlin's representation and the effective field theory (EFT) representation. We will show later that the
comparison between the results in these two representations leads to useful relations between the LECs in ChPT
and hadronic matrix elements calculable on lattice. To avoid discussing issues such as the gauge-dependence
of the LECs, throughout this paper we simply adopt the Feynman gauge which is the standard choice
in all papers of similar topics. 

\subsection{Sirlin's representation}

Earliest theory analysis of electromagnetic RCs in Fermi interactions can be traced back to the seminal
work by Kinoshita and Sirlin in 1958~\cite{Kinoshita:1958ru}, and later by Sirlin. He derived the universal
function $g(E,E_m,m)$ that summarizes the infrared (IR) physics of the RCs in generic beta decay
processes~\cite{Sirlin:1967zza}. The analysis was then extended to the full electroweak RCs, where the muon
decay rate was taken as a normalization~\cite{Sirlin:1974ni}. All these were later integrated into a
complete theory framework based on current algebra~\cite{Sirlin:1977sv} and the on-shell renormalization of
the SM electroweak sector~\cite{Sirlin:1980nh}, which we shall name as Sirlin's representation. Despite being
gradually superseded by the EFT representation, recently it was re-introduced in the study of $K_{l3}$ RCs
in a hybridized form with EFT, which aims to further reduce the existing theory uncertainty~\cite{Seng:2019lxf}.

In  Sirlin's representation, the $\mathcal{O}(G_F\alpha)$ electroweak RCs to the amplitude of a semi-leptonic
decay process of a spinless particle $\phi_i(p)\to \phi_f(p') e^+(p_e)\nu_e(p_\nu)$ can be summarized
as~\cite{Seng:2019lxf}:
\begin{eqnarray}
\delta M&=&\left[-\frac{\alpha}{2\pi}\left(\ln\frac{M_W	^2}{M_Z^2}+\frac{1}{4}\ln\frac{M_W^2}{m_e^2}
-\frac{1}{2}\ln\frac{m_e^2}{M_\gamma^2}+\frac{9}{8}+\frac{3}{4}a_\mathrm{pQCD}\right)+\frac{1}{2}
\delta_{\mathrm{HO}}^\mathrm{QED}\right]M_0\nonumber\\
&&-\frac{G_F}{\sqrt{2}}\bar{u}_\nu\gamma_\lambda(1-\gamma_5)v_e\delta F_{fi}^\lambda(p',p)+\delta M_{\gamma W}~.
\end{eqnarray}
The first line in the equation above represents the contributions from the ``weak'' RCs (see
Ref.~\cite{Seng:2019lxf} for rigorous definition) including its perturbative QCD (pQCD) corrections
$a_\mathrm{pQCD}\approx 0.068$, the electromagnetic RC to the electron wavefunction renormalization
(with a small photon mass $M_\gamma$ as an IR regulator), as well as the contribution from the resummation
of the large QED logs, which is formally of higher order but numerically sizable: $\delta_\mathrm{HO}^\mathrm{QED}
=0.0010(3)$~\cite{Erler:2002mv}. The second line encodes the contribution from the electromagnetic RCs to
the charged weak matrix element and the $\gamma W$ box diagram. Employing the on-mass-shell
formula~\cite{Brown:1970dd} and Ward identities, the form factor correction splits into two pieces:
$\delta F^\lambda_{fi}=\delta F^\lambda_{fi,2}+\delta F^\lambda_{fi,3}$, among which the ``two-point function''
contribution reads: 
\begin{equation}
\delta F_{fi,2}^\lambda(p',p)=-\frac{e	^2}{2}\int\frac{d^4q'}{(2\pi)^4}T^\mu_{fi\:\mu}(q';p',p)
\frac{\partial}{\partial q'_\lambda}\left(\frac{1}{q^{\prime 2}-M_\gamma^2}\frac{M_W^2}{M_W^2-q^{\prime 2}}\right),
\label{eq:2pt}
\end{equation}
where we have defined the ``generalized Compton tensor'' that consists of the interference between the
electromagnetic and charged weak current as:
\begin{equation}
T^{\mu\nu}_{fi}(q';p',p)=\int d^4xe^{iq'\cdot x}\left\langle \phi_f(p')\right|T\{J_\mathrm{em}^\mu(x)J_W^{\nu\dagger}(0)\}
\left|\phi_i(p)\right\rangle~.
\end{equation}
On the other hand, the explicit form of the ``three-point function'' contribution $\delta F_{fi,3}^\lambda$
is not of our concern. One needs only to know that it vanishes when the vector charged weak current
is conserved and $p-p'=0$. Finally, the $\gamma W$ box diagram contribution is given by:
\begin{equation}
\delta M_{\gamma W}=-\frac{G_Fe^2}{\sqrt{2}}\int\frac{d^4q'}{(2\pi)^4}\frac{\bar{u}_\nu\gamma^\nu(1-\gamma_5)
(\slashed{q}'-\slashed{p}_e+m_e)\gamma^\mu v_e}{(p_e-q')^2-m_e^2}\frac{1}{q^{\prime 2}-M_\gamma^2}
\frac{M_W^2}{M_W^2-q^{\prime 2}}T_{\mu\nu}^{fi}(q';p',p)~.\label{eq:gammaW}
\end{equation}
An important point to notice is that all the integrals above are ultraviolet (UV)-finite, so there is no
need to introduce any extra UV-regulators and unknown counterterms. 

Further simplifications can be made to the expressions above. First, using the on-shell formula
$(\slashed{p}_e+m_e)v_e=0$ and the Dirac matrix identity:
\begin{equation}
\gamma^\mu\gamma^\nu\gamma^\alpha=g^{\mu\nu}\gamma^\alpha-g^{\mu\alpha}\gamma^\nu+g^{\nu\alpha}\gamma^{\mu}
-i\epsilon^{\mu\nu\alpha\beta}\gamma_\beta\gamma_5,
\end{equation}
with $\epsilon^{0123}=-1$ in our convention, the lepton tensor in Eq.\eqref{eq:gammaW} can be rewritten as:
\begin{equation}
\bar{u}_\nu\gamma^\nu(1-\gamma_5)(\slashed{q}'-\slashed{p}_e+m_e)\gamma^\mu v_e=\bar{u}_\nu\gamma_\lambda
(1-\gamma_5)v_e\left[g^{\lambda\nu}q^{\prime\mu}+g^{\lambda\mu}q^{\prime\nu}-g^{\mu\nu}q^{\prime\lambda}-2g^{\lambda\nu}
p_e^{\mu}+i\epsilon^{\mu\nu\alpha\lambda}q_\alpha'\right]~.\label{eq:lepton}
\end{equation}
With this, the box diagram contribution in Eq.~\eqref{eq:gammaW} splits into two parts:
\begin{equation}
\delta M_{\gamma W}=\delta M_{\gamma W}^V+\delta M_{\gamma W}^A~,
\end{equation}
where $\delta M_{\gamma W}^V$ and $\delta M_{\gamma W}^A$ include the contribution from the first four
terms and the last term at the RHS of Eq.\eqref{eq:lepton}, respectively.

Next, we recall that the generalized Compton tensor satisfies the following Ward identities:
\begin{eqnarray}
q_\mu^{\prime}T^{\mu\nu}_{fi}(q';p',p)&=&-iF^\nu_{fi}(p',p)\nonumber\\
q_\nu T^{\mu\nu}_{fi}(q';p',p)&=&-iF^\mu_{fi}(p',p)-i\Gamma^\mu_{fi}(q';p',p)~,
\end{eqnarray}
where $q=p'+q'-p$, and
\begin{equation}
\Gamma^{\mu}_{fi}(q';p',p)=\int d^4xe^{iq'\cdot x}\left\langle \phi_f(p')\right|T\{J_\mathrm{em}^\mu(x)\partial\cdot
J_W^{\dagger}(0)\}\left|\phi_i(p)\right\rangle~.
\end{equation}
These Ward identities are derived from the equal-time commutation relation between the $J_W^{0\dagger}$ and
$J_\mathrm{em}^\mu$, i.e. the current algebra relation, which is protected from perturbative Quantum
Chromodynamics (pQCD) corrections to all orders.

With the identities above, the two-point function contribution (i.e. Eq.\eqref{eq:2pt}) and
$\delta M_{\gamma W}^V$ sums up to give:
\begin{eqnarray}
\delta M_2+\delta M_{\gamma W}^V&=&\frac{\alpha}{2\pi}\left[\ln\frac{M_W^2}{m_e^2}+\frac{3}{4}+\frac{1}{2}
\tilde{a}_g^\mathrm{res}\right]M_0+\frac{G_Fe^2}{\sqrt{2}}\bar{u}_\nu\gamma_\lambda(1-\gamma_5)v_e
\int\frac{d^4q'}{(2\pi)^4}\frac{M_W^2}{M_W^2-q^{\prime 2}}\nonumber\\
&&\times\frac{1}{(p_e-q')^2-m_e^2}\left\{\frac{2p_e\cdot q'q^{\prime\lambda}}{(q^{\prime 2}-M_\gamma^2)^2}
T^\mu_{fi\:\mu}(q';p',p)+\frac{2p_{e\mu}}{q^{\prime 2}-M_\gamma^2}T^{\mu\lambda}_{fi}(q';p',p)\right.\nonumber\\
&&\left.-\frac{(p-p')_\mu}{q^{\prime 2}-M_\gamma^2}T^{\lambda\mu}_{fi}(q';p',p)+\frac{i}{q^{\prime 2}-M_\gamma^2}
\Gamma^\lambda_{fi}(q';p',p)\right\}~.\label{eq:M2andMgammaWV}
\end{eqnarray}
Here, $\tilde{a}_g^\mathrm{res}\approx0.019$ is a small pQCD correction to the two-point function.
Using the free-field operator product expansion (OPE) of the hadronic tensors, it is easy to see that
the remaining integrals in the equation above do not depend on physics at the scale $q'\sim M_W$.

\subsection{\label{sec:EFT}The EFT representation}

The second and more commonly adopted representation in studies of the RCs in beta decays is based on the EFT
of the SM at low energy. In such a formalism, one constructs the most general Lagrangian consistent with
the symmetry properties of the underlying theory in terms of the relevant low-energy degrees of freedom (DOFs).
UV-divergences due to loop integrals are first regularized using dimensional regularization (DR) and then
canceled by the corresponding LECs. A power counting scheme is defined to ensure the finiteness of terms in
the Lagrangian for any given precision that one wants to achieve. Finally, a matching with the perturbative calculation in the SM at the UV-end is carried out to determine the dependence of the LECs on the UV-physics, e.g. large electroweak logarithms. 

For the decay processes we are discussing in this paper, i.e. $K_{l3}$ and $\pi_{e3}$, the corresponding
EFT is simply the three-flavor ChPT with dynamical photons and leptons. Here we shall simply quote the
involved chiral Lagrangian for future reference.
First, the pseudo-Nambu-Goldstone boson (pNGB) octet is contained in the usual matrix $U$. To describe its
coupling with the dynamical photon field $\mathcal{A}_\mu$, we introduce the following covariant derivative: 
\begin{equation}
D_\mu U=\partial_\mu U-i(r_\mu+q_R \mathcal{A}_\mu)U+iU(l_\mu+q_L \mathcal{A}_\mu)~,
\end{equation}
where we have introduced the left/right-handed external sources $\{l_\mu,r_\mu\}$ and spurion fields
$\{q_L,q_R\}$ that are traceless, Hermitian matrices in the quark flavor space. We also
define $u=\sqrt{U}$, and 
\begin{equation}
u_\mu=i[u^\dagger(\partial_\mu-ir_\mu-iq_R\mathcal{A}_\mu)u-u(\partial_\mu-il_\mu-iq_L\mathcal{A}_\mu)u^\dagger]~,
\end{equation}
as well as the covariant derivatives on the spurion fields:
\begin{equation}
\nabla_\mu q_R=\partial_\mu q_R-i[r_\mu,q_R],\:\:\:\nabla_\mu q_L=\partial_\mu q_L-i[l_\mu,q_L]~.
\end{equation}
Finally, for the SM charged weak interaction Lagrangian, the external sources should be identified as:
\begin{equation}
q_R=q_L=-eQ^\mathrm{em},\:\:l_\mu=\sum_l(\bar{l}\gamma_\mu \nu_{lL}Q_\mathrm{L}^\mathrm{w}+h.c.),\:\:r_\mu=0~,
\end{equation}
where 
\begin{equation}
Q^{\mathrm{em}}=\left(\begin{array}{ccc}
2/3 & 0 & 0\\
0 & -1/3 & 0\\
0 & 0 & -1/3
\end{array}\right),\:\:Q_{\mathrm{L}}^{\mathrm{w}}=-2\sqrt{2}G_{F}\left(\begin{array}{ccc}
0 & V_{ud} & V_{us}\\
0 & 0 & 0\\
0 & 0 & 0
\end{array}\right)~.
\end{equation}
One sees that the dynamical leptons enter through the left-handed source field $l_\mu$.

Now we can write down the chiral Lagrangian. In a consistent chiral power counting scheme, $p$
(a typical small momentum of the pNGBs) and $e$ should carry the same chiral order. Therefore
at leading order (LO) we have:
\begin{eqnarray}
\mathcal{L}^{(2)}&=&\frac{F_0^2}{4}\left\langle D_\mu U(D^\mu U)^\dagger+U\chi^\dagger+\chi U^\dagger\right\rangle
+ZF_0^4 \left\langle q_L U^\dagger q_R U\right\rangle-\frac{1}{4}F_{\mu\nu}F^{\mu\nu}-\frac{1}{2\xi}
(\partial_\mu \mathcal{A}^\mu)^2\nonumber\\
&&+\frac{1}{2}M_\gamma^2 \mathcal{A}_\mu \mathcal{A}^\mu+\sum_l[\bar{l}(i\slashed{\partial}
+e\slashed{\mathcal{A}}-m_l)l+\bar{\nu}_{lL}i\slashed{\partial}\nu_{iL}]~,
\end{eqnarray}
where $F_0$ is the pion decay constant in the chiral limit, $F_{\mu\nu}$ is the photon field strength tensor,
$\chi=2B_0M_q$ with $M_q$ the quark mass matrix, and $Z\approx 0.8$ is obtained from the $\pi^\pm-\pi^0$ mass
splitting. The notation $\left\langle...\right\rangle$ represents the trace over the flavor space.
As stated above, throughout this work we choose $\xi=1$, the Feynman gauge.

To absorb the UV-divergences generated from $\mathcal{L}^{(2)}$ at one loop, one needs to introduce the
next-to-leading order (NLO) chiral Lagrangian, which could either scale as $\mathcal{O}(p^4)$ or
$\mathcal{O}(e^2p^2)$. The former is just the standard Gasser-Leutwyler Lagrangian~\cite{Gasser:1984gg}
so we shall concentrate on the latter.
There are two types of chiral Lagrangian at $\mathcal{O}(e^2p^2)$. The first type characterizes the
short-distance electromagnetic effects of hadrons~\cite{Urech:1994hd,Neufeld:1995mu}:
\begin{eqnarray}
\mathcal{L}^{e^2p^2}_{\{K\}}&=&F_0^2\left\{\frac{1}{2}K_1\left\langle D^\mu U(D_\mu U)^\dagger\right\rangle\left\langle q_Rq_R+q_Lq_L\right\rangle+K_2\left\langle D^\mu U(D_\mu U)^\dagger\right\rangle\left\langle q_R Uq_L U^\dagger\right\rangle\right.\nonumber\\
&&+K_3\left(\left\langle(D^\mu U)^\dagger q_R U\right\rangle\left\langle(D_\mu U)^\dagger q_R U\right\rangle+\left\langle D^\mu Uq_L U^\dagger\right\rangle\left\langle D_\mu Uq_L U^\dagger\right\rangle\right)\nonumber\\
&&K_4\left\langle(D^\mu U)^\dagger q_R U\right\rangle\left\langle D_\mu Uq_L U^\dagger\right\rangle+K_5\left\langle q_Lq_L(D^\mu U)^\dagger D_\mu U+q_Rq_RD^\mu U(D_\mu U)^\dagger\right\rangle\nonumber\\
&&+K_6\left\langle(D^\mu U)^\dagger D_\mu Uq_L U^\dagger q_R U+D^\mu U(D_\mu U)^\dagger q_R Uq_LU^\dagger\right\rangle\nonumber\\
&&+\frac{1}{2}K_7\left\langle\chi^\dagger U+U^\dagger\chi\right\rangle\left\langle q_Rq_R+q_Lq_L\right\rangle+K_8\left\langle\chi^\dagger U+U^\dagger \chi\right\rangle\left\langle q_RUq_LU^\dagger\right\rangle\nonumber\\
&&+K_9\left\langle(\chi^\dagger U+U^\dagger \chi)q_Lq_L+(\chi U^\dagger+U\chi^\dagger)q_Rq_R\right\rangle\nonumber\\
&&+K_{10}\left\langle(\chi^\dagger U+U^\dagger\chi)q_L U^\dagger q_R U+(\chi U^\dagger+U\chi^\dagger)q_R Uq_LU^\dagger\right\rangle\nonumber\\
&&+K_{11}\left\langle(\chi^\dagger U-U^\dagger \chi)q_LU^\dagger q_R U+(\chi U^\dagger-U\chi^\dagger)q_R Uq_LU^\dagger\right\rangle\nonumber\\
&&+K_{12}\left\langle(D^\mu U)^\dagger[\nabla_\mu q_R,q_R]U+D^\mu U[\nabla_\mu q_L,q_L]U^\dagger\right\rangle\nonumber\\
&&\bigl.+K_{13}\left\langle\nabla^\mu q_R U\nabla_\mu q_L U^\dagger\right\rangle+K_{14}\left\langle\nabla^\mu q_R\nabla_\mu q_R+\nabla^\mu q_L\nabla_\mu q_L\right\rangle\biggr\},\label{eq:LKi}
\end{eqnarray}
although the lepton fields may still enter through the covariant derivatives. The second type involves
explicit leptonic degrees of freedom. The part relevant to $K_{l3}$ and $\pi_{e3}$ RCs is
given by~\cite{Knecht:1999ag}:
\begin{eqnarray}
\mathcal{L}^{e^2p^2 }_{\{X\}}&=&e^2F_0^2\sum_l\left\{X_1\bar{l}\gamma_\mu \nu_{lL}\left\langle u^\mu\left\{\mathcal{Q}_\mathrm{R}^\mathrm{em},\mathcal{Q}_\mathrm{L}^\mathrm{w}\right\}\right\rangle+X_2\bar{l}\gamma_\mu \nu_{lL}\left\langle u^\mu\left[\mathcal{Q}_\mathrm{R}^\mathrm{em},\mathcal{Q}_\mathrm{L}^\mathrm{w}\right]\right\rangle\right.\nonumber\\
&&\left.+X_3m_l\bar{l}v_{lL}\left\langle\mathcal{Q}_\mathrm{L}^\mathrm{w}\mathcal{Q}^\mathrm{em}_\mathrm{R}\right\rangle+h.c.\right\}+e^2\sum_l X_6\bar{l}(i\slashed{\partial}+e\slashed{\mathcal{A}})l~,
\end{eqnarray}
where $\mathcal{Q}_\mathrm{R}^\mathrm{em}=u^\dagger Q^\mathrm{em}u$ and $\mathcal{Q}_\mathrm{L}^\mathrm{w}=
uQ_\mathrm{L}^\mathrm{w}u^\dagger$. 

The LECs $\{K_i,X_i\}$ are generically UV-divergent, and their corresponding renormalized LECs are defined as: 
\begin{equation}
K_i^r(\mu)=K_i-\Sigma_i\lambda,\:\:\:X_i^r(\mu)=X_i-\Xi_i\lambda,
\end{equation}
where 
\begin{equation}
\lambda=\frac{\mu^{d-4}}{16\pi^2}\left(\frac{1}{d-4}-\frac{1}{2}\left[\ln 4\pi-\gamma_E+1\right]\right),
\end{equation}
with $\mu$ the scale introduced in DR, $d$ the number of the space-time dimensions, and $\gamma_E$
the Euler-Mascheroni constant. The values of $\{\Sigma_i,\Xi_i\}$ are given in
Refs.~\cite{Urech:1994hd,Knecht:1999ag}, respectively. In connection with the SM electroweak sector, we find
that $X_6^r$ and $K_{12}^r$ are sensitive to physics at the scale $q\sim M_W$ (in another word, they
carry the large electroweak logarithms). It is customary to define the combination $X_6^\mathrm{phys}(\mu)
\equiv X_6^r(\mu)-4K_{12}^r(\mu)$ and take $\mu=M_\rho$ in the numerical analysis.

\begin{table}
	\begin{centering}
		\begin{tabular}{|c|c|}
			\hline 
			& $\delta_{\mathrm{em}}^{Kl}$(\%)\tabularnewline
			\hline 
			\hline 
			$K_{e3}^{0}$ & $0.99\pm0.19_{e^{2}p^{4}}\pm0.11_{\mathrm{LEC}}$\tabularnewline
			\hline 
			$K_{e3}^{\pm}$ & $0.10\pm0.19_{e^{2}p^{4}}\pm0.16_{\mathrm{LEC}}$\tabularnewline
			\hline 
			$K_{\mu3}^{0}$ & $1.40\pm0.19_{e^{2}p^{4}}\pm0.11_{\mathrm{LEC}}$\tabularnewline
			\hline 
			$K_{\mu3}^{\pm}$ & $0.016\pm0.19_{e^{2}p^{4}}\pm0.16_{\mathrm{LEC}}$\tabularnewline
			\hline 
		\end{tabular}
		\par\end{centering}
	\caption{\label{tab:deltaem}$\delta_{\mathrm{em}}^{Kl}$ calculated in ChPT~\cite{Cirigliano:2008wn}.}
\end{table}

With the effective Lagrangian above, the RCs to $K_{l3}$ and $\pi_{e3}$ were computed to
$\mathcal{O}(e^2p^2)$~\cite{Cirigliano:2001mk,Cirigliano:2002ng,Cirigliano:2008wn}, and we shall briefly
discuss the main results. First, the master formula of the $K_{l3}$ decay rate is given by:
\begin{equation}
\Gamma_{K_{l3}}=\frac{C_K^2G_F^2M_K^5}{128\pi^3}S_\mathrm{EW}|F_+^{\pi^-K^0}(0)|^2I_{Kl}^{(0)}(\lambda_i)
\left(1+\delta_\mathrm{em}^{Kl}+\delta_\mathrm{SU(2)}^{K\pi}\right)~,
\end{equation}
among which the short-distance electroweak factor $S_\mathrm{EW}$ is defined as\footnote{There is a typo in
Eq.~(94) of Ref.~\cite{DescotesGenon:2005pw}, the factor 1/2 in front of $e^2$ should not be there.}:
\begin{equation}
S_\mathrm{EW}\equiv 1-e^2\left[-\frac{1}{2\pi^2}\ln\frac{M_Z}{M_\rho}+(X_6^\mathrm{phys})_{\alpha_s}\right]
+\delta_\mathrm{HO}^\mathrm{QED}=1.0229(3)~,
\end{equation}
where we take $M_\rho=0.77$~GeV. Here $(X_6)^\mathrm{phys}_{\alpha_s}\approx 3.0\times 10^{-3}$~\cite{DescotesGenon:2005pw}
summarizes the $\mathcal{O}(\alpha_s)$ pQCD contribution to $X_6^\mathrm{phys}$ (but not from higher-order
contributions such as $\mathcal{O}(\alpha_s^{(2)})$, which we shall discuss later). This value is consistent
with that quoted in Ref.~\cite{Cirigliano:2003yr} as well as the more commonly cited value of 1.0232 by Marciano
and Sirlin~\cite{Marciano:1993sh}\footnote{On the other hand, the quoted value of $S_\mathrm{EW}=1.0223(5)$
in Ref.~\cite{Cirigliano:2011ny} was inconsistent with the subsequent $V_{us}$ phenomenology in the same paper,
and therefore should not be used.}.
Meanwhile, the long-distance EM correction is represented by the quantity $\delta_\mathrm{em}^{Kl}$. The ChPT
estimations of their numerical values in different channels are summarized in Tab.~\ref{tab:deltaem}. We see
that there are two sources of uncertainties in $\delta_\mathrm{em}^{Kl}$, namely (1) the neglected higher-order
terms in the chiral power counting, and (2) the LECs $\{K_i^r,X_i^r\}$. 
Here we are only interested in its dependence on the non-unsuppressed  LECs (i.e. those contributing to
$\delta F_+^{\pi K}$)\footnote{Notice that $X_1$ is scale-independent, so $X_1^r=X_1$. The same goes
for $K_7$, $K_{13}$ and $K_{14}$ in the Feynman gauge.}:
\begin{eqnarray}
\delta_\mathrm{em}^{K^\pm l}&=&2e^2\left[-\frac{8}{3}X_1-\frac{1}{2}\tilde{X}_6^\mathrm{phys}(M_\rho)-2K_3^r(M_\rho)
 +K_4^r(M_\rho)+\frac{2}{3}K_5^r(M_\rho)+\frac{2}{3}K_6^r(M_\rho)\right]+... ,\nonumber\\
\delta_\mathrm{em}^{K^0 l}&=&2e^2\left[\frac{4}{3}X_1-\frac{1}{2}\tilde{X}_6^\mathrm{phys}(M_\rho)\right]+... ,
\label{eq:deltaemKl3}
\end{eqnarray} 
where $\tilde{X}_6^\mathrm{phys}(M_\rho)\equiv X_6^\mathrm{phys}(M_\rho)+(2\pi^2)^{-1}\ln(M_Z/M_\rho)
-(X_6^\mathrm{phys})_{\alpha_s}$ removes the large electroweak logarithm and the $\mathcal{O}(\alpha_s)$
pQCD correction from $X_6^\mathrm{phys}$.
As a comparison, we can define a similar quantity for $\pi_{e3}$, and its LEC-dependence reads:
\begin{equation}
\delta_\mathrm{em}^{\pi^\pm e}=2e^2\left[-\frac{2}{3}X_1-\frac{1}{2}\tilde{X}_6^\mathrm{phys}(M_\rho)\right]+... \ .
\label{eq:deltaempie3}
\end{equation}

It is useful to contrast the results above with the case of the kaon and pion leptonic beta decay.
We notice that both the $K_{l2}$ and $\pi_{l2}$ decay rate depend on the same combination of LECs~\cite{Knecht:1999ag}:
\begin{equation}
E^r\equiv\frac{8}{3}K_1^r+\frac{8}{3}K_2^r+\frac{20}{9}K_5^r+\frac{20}{9}K_6^r-\frac{4}{3}X_1-4X_2^r+4X_3^r
-X_6^\mathrm{phys},
\end{equation}
so it will be canceled out in the ratio $R_A=\Gamma_{K_{\mu2}}/\Gamma_{\pi_{\mu2}}$, which results in a reduced
theory uncertainty in the extraction of the ratio $|V_{us}/V_{ud}|$. This is, however, not the case in the
ratio $R_V=\Gamma_{K_{l3}}/\Gamma_{\pi_{e3}}$ recently introduced in Ref.~\cite{Czarnecki:2019iwz}, as we see
that Eqs.~\eqref{eq:deltaemKl3} and \eqref{eq:deltaempie3} are not identical (except the $\tilde{X}_6^\mathrm{phys}$
term which is common to all channels). Therefore, to reduce the theoretical uncertainty in $R_V$ we propose
a first-principles calculation of $X_1$ and $-2K_3^r+K_4^r+(2/3)(K_5^r+K_6^r)$ and outline an appropriate method below.

\section{\label{sec:axialbox}Lattice QCD calculation of $X_1$ and $X_6^\mathrm{phys}$ via the
  $\gamma W$ box}

We start by discussing the LECs $X_1$ and $X_6^\mathrm{phys}$. They describe the electromagnetic interaction
between leptons and pNGBs, so it is natural to expect that they could be related to the hadronic matrix
element that occurs in the $\gamma W$ box diagram, Eq.~\eqref{eq:gammaW}. This section serves to derive
such a relation.

We first consider the electroweak RCs in the decay process $\phi_i\rightarrow \phi_f e^+\nu_e$ in Sirlin's
representation, and restrict ourselves to the case where $M_{\phi_i}\approx M_{\phi_f}\gg m_e$. In this limit,
we can define a power counting where $p-p'$, $p_e$ and $p_\nu$ all scale as a small expansion parameter
$\Delta$. An enormous amount of simplification is observed if we retain the terms in $\delta M$ only
up to $\mathcal{O}(\Delta^0)$:
\begin{enumerate}
	\item The three-point function contribution to $\delta F_{fi}^\mu$ vanishes;
	\item The weak axial charged-current contribution to the integrals in Eq.~\eqref{eq:M2andMgammaWV}
          vanishes. The vector contribution does not vanish, but it survives only in the region where
          $q'\sim \Delta$, so it is sufficient to replace $T^{\mu\nu}_{fi}$ and $\Gamma_{fi}^\mu$ by their
          respective ``convection terms''~\cite{Meister:1963zz} that describe the IR behavior of these
          quantities. By doing so, the integrals in Eq.~\eqref{eq:M2andMgammaWV} are analytically calculable.
	\item The remainder of the $\gamma W$ box contribution simplifies to 
              $\delta M_{\gamma W}^A=\Box_{\gamma W}^{VA}(\phi_f,\phi_i)M_0$,
              where
\begin{equation}
\Box_{\gamma W}^{VA}(\phi_f,\phi_i)\equiv\frac{ie^2}{2M_{\phi_i}^2}\int\frac{d^4q}{(2\pi)^4}\frac{1}{(q^2)^2}
\frac{M_W^2}{M_W^2-q^2}\epsilon^{\mu\nu\alpha\beta}q_\alpha p_\beta \frac{T_{\mu\nu}^{fi}(q;p,p)}{F_+^{fi}(0)}\label{eq:box}
\end{equation} 
shall be denoted as the ``forward axial $\gamma W$ box'', as it probes the axial charged weak current
in $T_{\mu\nu}^{fi}$. 
\end{enumerate}

From the above, we see that in the $\Delta\to 0$ limit the only unknown
piece in $\delta M$ is $\Box_{\gamma W}^{VA}(\phi_f,\phi_i)$ which depends on
the details of the non-perturbative QCD at the hadron scale. It is, however, a
well-defined hadronic matrix element which is calculable on
lattice. In fact, Ref.\cite{Feng:2020zdc} presented a first-principles calculation of
$\Box_{\gamma W}^{VA}(\pi^0,\pi^+)$ by combining the direct computation of the relevant four-point
contraction diagrams at small $Q^2=-q^2$ and a pQCD calculation to $\mathcal{O}(\alpha_s^4)$ at large
$Q^2$, achieving an impressive 1\% overall accuracy. Other possible methods include the application of
the Feynman-Hellmann theorem on lattice \cite{Seng:2019plg,Bouchard:2016heu,Chambers:2017dov}.

Now it is clear how one could obtained the LECs $X_1$ and $X_6^\mathrm{phys}$ on the lattice: One repeats
the calculation of $\delta M$ in the ChPT and take the $\Delta\to 0$ limit, in which the only quantities
that are not  determined \textit{a priori} are the LECs. Therefore, comparing the expression of $\delta M$
in the $\Delta\to 0$ limit between Sirlin's representation and the ChPT representation gives us a
relation between $\{X_1,X_6^\mathrm{phys}\}$ and $\Box_{\gamma W}^{VA}$\footnote{See Ref.\cite{Passera:2011ae} for an early attempt to compare these two representations.}. Of course, one needs to calculate
the latter at least in two different channels to fix $X_1$ and $X_6^\mathrm{phys}$ individually. In
what follows we choose $\pi_{e3}$ and $K_{e3}^0$ to fulfill this task.

\subsection{Axial $\gamma W$ box diagram in $\pi_{e3}$ decay}

In the $\pi_{e3}$ channel, since the strong isospin breaking effects are small, the $\Delta\to 0$ limit is
in fact quite well-satisfied in nature (the same holds for the free neutron and nuclear beta decays). To
evaluate the integrals in Eq.~\eqref{eq:M2andMgammaWV}, we replace $T^{\mu\nu}$ and $\Gamma^\mu$ by their
convection terms:
\begin{eqnarray}
T_{\pi^0\pi^+}^{\mu\nu}(q';p',p)&\rightarrow& \frac{i(2p-q')^\mu F_{\pi^0\pi^+}^\nu(p',p)}{(p-q')^2-M_\pi^2}\nonumber\\
\Gamma_{\pi^0\pi^+}^\mu(q';p',p)&\rightarrow&-\frac{(2p-q')^\mu (p'-p)\cdot F_{\pi^0\pi^+}(p',p)}{(p-q')^2-M_\pi^2}~.
\end{eqnarray}
With these, the total one-loop electroweak RCs to the decay amplitude in Sirlin's representation read
($u=(p-p_e)^2$, $\beta=|\vec{p}_e|/E_e$):
\begin{eqnarray}
\delta M&=&M_0\left\{\frac{\alpha}{4\pi}\left[\frac{3}{2}\ln\frac{M_W^2}{m_e^2}-2\ln\frac{M_W^2}{M_Z^2}+2
\ln\frac{m_e^2}{M_\gamma^2}-\frac{11}{4}+\tilde{a}_g+4p_e\cdot pC_0(u,M_\pi,m_e)+\frac{1}{\beta}
\ln\frac{1+\beta}{1-\beta}\right]\right.\nonumber\\
&&\left.+\Box_{\gamma W}^{VA}(\pi_0,\pi_+)+\frac{1}{2}\delta_\mathrm{HO}^\mathrm{QED}\right\}+\frac{\alpha}{4\pi}
\frac{G_F}{\sqrt{2}}\bar{u}_\nu\slashed{p}_e(1-\gamma_5)v_e\frac{p\cdot F_{\pi^0\pi^+}}{p\cdot p_e}
\frac{1}{\beta}\ln\frac{1+\beta}{1-\beta}+\mathcal{O}(\Delta)~.\label{eq:pie3Sirlin}
\end{eqnarray}
Here, $\tilde{a}_g=-(3/2)a_\mathrm{pQCD}+\tilde{a}_g^\mathrm{res}\approx -0.083$ summarizes the $\mathcal{O}(\alpha_s)$
pQCD correction to all one-loop diagrams except the axial $\gamma W$ box\footnote{This pQCD correction is
small because it is not attached to a large electroweak logarithm, so it is not necessary to include terms
with higher powers in $\alpha_s$. In fact this term is usually discarded in most papers. Here we retain it
for completeness.}. Meanwhile, $C_0$ is the well-known IR-divergent loop function:
\begin{equation}
C_0(z,m_1,m_2)=\int\frac{d^4q}{i\pi^2}\frac{1}{(q^2-M_\gamma^2+i\epsilon)((q+p_1)^2-m_1^2+i\epsilon)
    ((q-p_2)^2-m_2^2+i\epsilon)}~,
\end{equation}
with $p_1^2=m_1^2$, $p_2^2=m_2^2$ and $z=(p_1+p_2)^2$.
On the other hand, taking the $\Delta\to 0$ limit in the $\mathcal{O}(e^2p^2)$ ChPT
expression~\cite{Cirigliano:2002ng} gives:
\begin{eqnarray}
\delta M&=&M_0\left\{\frac{\alpha}{4\pi}\left[-\frac{3}{2}-\frac{3}{2}\ln\frac{m_e^2}{\mu^2}
+2\ln\frac{m_e^2}{M_\gamma^2}+4p_e\cdot pC_0(u,M_\pi,m_e)+\frac{1}{\beta}\ln\frac{1+\beta}{1-\beta}\right]
+\frac{1}{2}\delta_\mathrm{HO}^\mathrm{QED}\right.\nonumber\\
&&\left.+e^2\left(-\frac{2}{3}X_1-\frac{1}{2}X_6^\mathrm{phys}\right)\right\}+\frac{\alpha}{4\pi}
\frac{G_F}{\sqrt{2}}\bar{u}_\nu\slashed{p}_e(1-\gamma_5)v_e\frac{p\cdot F_{\pi^0\pi^+}}{p\cdot p_e}
\frac{1}{\beta}\ln\frac{1+\beta}{1-\beta}+\mathcal{O}(\Delta)~.\nonumber\\\label{eq:pie3ChPT}
\end{eqnarray}
We see that Eq.~\eqref{eq:pie3Sirlin} and \eqref{eq:pie3ChPT} agree completely in their IR behavior,
which is of course expected. 

We now want to equate these two expressions to obtain the relation between $X_i$ and $\Box_{\gamma W}^{VA}$.
In doing so, we find the definition of $\tilde{X}_6^\mathrm{phys}$ to be not particularly convenient,
because (1) in Ref.\cite{Feng:2020zdc} the pQCD correction is evaluated up to $\mathcal{O}(\alpha_s^4)$
instead of just $\mathcal{O}(\alpha_s)$, and (2) in the first-principles evaluation of Eq.~\eqref{eq:box},
one requires a smooth connection between the pQCD-corrected integrand in the asymptotic region and
the non-perturbative integrand at small $Q^2$. Thus, the procedure to ``remove the pQCD correction''
becomes rather unnatural. Therefore, we choose instead to express our result in terms of
\begin{equation}
\bar{X}_6^\mathrm{phys}(M_\rho)\equiv X_6^\mathrm{phys}(M_\rho)+\frac{1}{2\pi^2}\ln\frac{M_Z}{M_\rho}~,
\label{eq:X6bar}
\end{equation}
that removes only the large electroweak logarithm but retains the full pQCD corrections to all orders.
With this we obtain
\begin{equation}
\frac{4}{3}X_1+\bar{X}_6^\mathrm{phys}(M_\rho)=-\frac{1}{2\pi\alpha}\left(\Box_{\gamma W}^{VA}(\pi_0,\pi_+)
-\frac{\alpha}{8\pi}\ln\frac{M_W^2}{M_\rho^2}\right)+\frac{1}{8\pi^2}\left(\frac{5}{4}-\tilde{a}_g\right)~,
\label{eq:pie3matching}
\end{equation}
which is the first central result in this paper: It matches a specific linear combination of $X_1$ and
$\bar{X}_6^\mathrm{phys}$ to the axial $\gamma W$ box in $\pi_{e3}$ decay. We observe that in the first bracket
at the right of Eq.~\eqref{eq:pie3matching}, the large electroweak logarithm contribution to $\Box_{\gamma W}^{VA}$
has been subtracted out due to the use of $\bar{X}_6^\mathrm{phys}$ at the left.

Substituting the lattice QCD result $\Box_{\gamma W}^{VA}(\pi_0,\pi_+)=2.830(28)\times 10^{-3}$~\cite{Feng:2020zdc}
gives:
\begin{equation}
\frac{4}{3}X_1+\bar{X}_6^\mathrm{phys}(M_\rho)=0.0140(6)_\mathrm{box}(8)_\mathrm{ChPT}~,
\end{equation}
where the first uncertainty comes from the box diagram, and the second is the estimated leading ChPT
uncertainty that comes from the neglected $\pi^0-\eta$ mixing terms which scale as $M_\pi^2/(M_\eta^2-M_\pi^2)
\sim 6$\%.  It is instructive to compare the result above with that from the resonance
model~\cite{DescotesGenon:2005pw}. There, they estimated $X_1=-3.7\times 10^{-3}$ and
$\bar{X}_6^\mathrm{phys}=\tilde{X}_6^\mathrm{phys}+(X_6^\mathrm{phys})_{\alpha_s}=(10.4+3.0)\times 10^{-3}$,
with no robust estimation of the theory uncertainty. That implies
\begin{equation}
\frac{4}{3}X_1+\bar{X}_6^\mathrm{phys}(M_\rho)=0.0085,\:\:\:(\mathrm{resonance}\:\mathrm{model})
\end{equation}
which is significantly below the lattice result. This suggests that a careful first-principles
study of the LECs could lead to a visible change in the central values of  $\delta_\mathrm{em}$.

\subsection{Axial $\gamma W$ box diagram in $K_{e3}^0$ deacy}

The same matching can in principle also be done on $K_{e3}^0$ deacy in order to determine another
linear combination of $X_1$ and $\bar{X}_6^\mathrm{phys}$. The only extra complication is that $M_K$ is
significantly larger than $M_\pi$ so the $\Delta\to 0$ limit is not satisfied in nature. Nevertheless,
nothing prohibits us from considering an unphysical situation where $M_K\approx M_\pi\equiv M_\phi$,
which is always achievable on the lattice, the well-known SU(3) limit. In this limit all the
simplifications in  Sirlin's representation work again, provided that the axial $\gamma W$ box diagram for
$K_{e3}^0$ decay is now evaluated at the SU(3) symmetric point (i.e. $m_u=m_d=m_s$) rather than on the physical
point.  Despite such an unphysical setting, the LECs extracted from this procedure can still be applied to
physical processes because they are by definition independent of the quark masses.  

To evaluate the integrals in Eq.~\eqref{eq:M2andMgammaWV}, one again replaces $T^{\mu\nu}$ and $\Gamma^\mu$
by their convection terms. In this case they read:
\begin{eqnarray}
T_{\pi^-K^0}^{\mu\nu}(q';p',p)&\rightarrow &-\frac{i(2p'+q')^\mu F_{\pi^-K^0}^\nu(p',p)}{(p'+q')^2-M_\phi^2}~,\nonumber\\
\Gamma^\mu_{\pi^-K^0}(q';p',p)&\rightarrow & \frac{(2p'+q')^\mu (p'-p)\cdot F_{\pi^-K^0}(p',p)}{(p'+q')^2-M_\phi^2}~.
\end{eqnarray}
With these, the total one-loop electroweak RCs to the $K_{e3}^0$ decay amplitude in Sirlin's representation
with the unphysical setting reads ($s=(p'+p_e)^2$, $\beta=|\vec{p}_e|/E_e$)\footnote{We take this opportunity
to point out that the definition of the quantity $X$ in Eq.(B.1) of Ref.\cite{Cirigliano:2001mk} is
incorrect. The correct definition follows Eq.(2.7) in Ref.\cite{Beenakker:1988jr}.}:
\begin{eqnarray}
\delta M&=&M_0\left\{\frac{\alpha}{4\pi}\left[\frac{3}{2}\ln\frac{M_W^2}{m_e^2}-2\ln\frac{M_W^2}{M_Z^2}
+2\ln\frac{m_e^2}{M_\gamma^2}-\frac{11}{4}+\tilde{a}_g-4p_e\cdot p'C_0(s,M_\phi,m_e)\right.\right.\nonumber\\
&&\left.\left.-\frac{2}{\beta}\ln\left(-\sqrt{\frac{1-\beta}{1+\beta}}+i\epsilon\right)\right]
+\left(\Box_{\gamma W}^{VA}(\pi^-,K^0)\right)_\mathrm{SU(3)}+\frac{1}{2}\delta_\mathrm{HO}^\mathrm{QED}\right\}\nonumber\\
&&-\frac{\alpha}{2\pi}\frac{G_F}{\sqrt{2}}\bar{u}_\nu\slashed{p}_e(1-\gamma_5)v_e\frac{p'\cdot
F_{\pi^-K^0}}{p'\cdot p_e}\frac{1}{\beta}\ln\left(-\sqrt{\frac{1-\beta}{1+\beta}}+i\epsilon\right)+\mathcal{O}(\Delta)~.
\end{eqnarray}
Here, the subscript in $\left(\Box_{\gamma W}^{VA}(\pi^-,K^0)\right)_\mathrm{SU(3)}$ reminds us that it should
be evaluated at the SU(3) symmetric point. On the other hand, in the $\Delta\to 0$ limit the ChPT
expression~\cite{Cirigliano:2001mk} reads:
\begin{eqnarray}
\delta M&=&M_0\left\{\frac{\alpha}{4\pi}\left[-\frac{3}{2}-\frac{3}{2}\ln\frac{m_e^2}{\mu^2}
+2\ln\frac{m_e^2}{M_\gamma^2}-4p_e\cdot p'C_0(s,M_\phi,m_e)\right.\right.\nonumber\\
&&\left.\left.-\frac{2}{\beta}\ln\left(-\sqrt{\frac{1-\beta}{1+\beta}}+i\epsilon\right)\right]
+\frac{1}{2}\delta_\mathrm{HO}^\mathrm{QED}+e^2\left(\frac{4}{3}X_1-\frac{1}{2}X_6^\mathrm{phys}\right)\right\}\nonumber\\
&&-\frac{\alpha}{2\pi}\frac{G_F}{\sqrt{2}}\bar{u}_\nu\slashed{p}_e(1-\gamma_5)v_e\frac{p'\cdot
F_{\pi^-K^0}}{p'\cdot p_e}\frac{1}{\beta}\ln\left(-\sqrt{\frac{1-\beta}{1+\beta}}+i\epsilon\right)+\mathcal{O}(\Delta)~.
\end{eqnarray}
Matching the two expressions gives: 
\begin{equation}
-\frac{8}{3}X_1+\bar{X}_6^\mathrm{phys}(M_\rho)=-\frac{1}{2\pi\alpha}\left(\left(\Box_{\gamma W}^{VA}(\pi_-,K_0)
\right)_\mathrm{SU(3)}-\frac{\alpha}{8\pi}\ln\frac{M_W^2}{M_\rho^2}\right)+\frac{1}{8\pi^2}
\left(\frac{5}{4}-\tilde{a}_g\right)~,\label{eq:Ke30matching}
\end{equation}
which is the second central result in this paper.
Therefore, a future lattice calculation of $\left(\Box_{\gamma W}^{VA}(\pi_-,K_0)\right)_\mathrm{SU(3)}$ allows
a simultaneous determination of $X_1$ and $\bar{X}_6^\mathrm{phys}(M_\rho)$ from first principles. A point
to remember is that the matching above is valid only up to $\mathcal{O}(e^2p^2)$, therefore taking
$M_\phi^2\ll \Lambda_\chi$ in the lattice calculation will help suppressing the theory uncertainties
from the neglected $\mathcal{O}(e^2p^4)$ terms.
In the flavor SU(3) limit, the $K^0_{e3}$ $\gamma$W-box diagrams share the same types of quark
contractions as $\pi^0_{e3}$ in the lattice calculation. Therefore, it is straightforward to extend the
calculation of $\gamma$W-box diagrams from the pion to the kaon sector.

One may wonder if calculating the axial $\gamma W$ box diagrams in more channels, such as $K_{e3}^+$, will
also give us information about other LECs, for example the $\{K_i^r\}$ that appear in $\delta_\mathrm{em}^{K^\pm l}$
(see Eq.~\eqref{eq:deltaemKl3}). This is, unfortunately, impossible because through a simple inspection of
Eq.~\eqref{eq:LKi} one sees that the terms with these LECs can survive even without a lepton, which
means that they do not describe a short-distance lepton-hadron QED interaction, hence the axial
$\gamma W$ box cannot carry any information of these LECs. To study them, we must construct another
type of correlation functions calculable on lattice, which we shall discuss in the following section.

\section{\label{sec:4point}The setup of a lattice QCD calculation of the $\{K_i^r\}$ }

As far as the unsuppressed contribution to the $K_{l3}$ decay rate is concerned, the only extra
LEC we need to calculate is the combination $-2K_3^r+K_4^r+(2/3)(K_5^r+K_6^r)$ (see Eq.\eqref{eq:deltaemKl3}).
However, if we wish to be more precise by also studying the RCs to the form factor $F_-^{\pi K}$, then we
need to know $K_3^r,...,K_6^r$ individually~\cite{Cirigliano:2001mk}. At the same time, $K_1^r$ and $K_2^r$
are also interesting because in the large-$N_c$ limit they satisfy the relations $K_3^r=-K_1^r$ and
$K_4^r=2K_2^r$, \cite{Urech:1994hd,Bijnens:1996kk}, so by calculating them one could test the precision
of the large-$N_c$ predictions from first principles. Therefore in this section we shall outline
a strategy to calculate $K_1^r,...,K_6^r$ on the lattice. While the remaining $\{K_i^r\}$ are also
interesting by themselves (e.g., $K_8^r,...,K_{11}^r$ contribute to the $K^\pm-K^0$ mass splitting
at $\mathcal{O}(e^2p^2)$~\cite{Urech:1994hd,Moussallam:1997xx}), we will not discuss them here.

Ref.~\cite{Ananthanarayan:2004qk} expressed the $\{K_i^r\}$ in terms of a series of four-point functions,
which they later calculated using resonance models to obtain an estimate of the LECs. We find that such
a formalism is indeed a good starting point to motivate a realistic lattice QCD calculation upon
appropriate modifications (for instance, the chiral limit, which is not attainable on lattice).
In what follows, we shall derive the modified four-point function representation of the LECs. Of course we could work on the physical point, but since the variation of non-zero quark masses do not give rise to extra singularities in these correlation functions (which can be seen from the Feynman diagrams in Fig.\ref{fig:mesonloop}, \ref{fig:photonWOpole} and \ref{fig:photonWpole}), here we shall present our result in the SU(3) limit, $M_\pi=M_K=M_\eta\equiv M_\phi$, which brings a great simplification to the involved loop functions.

\subsection{Lepton-free Lagrangian with external sources and spurions}

We start again by discussing the SM Lagrangian responsible for the semi-leptonic beta decay processes,
which was explained in some detail in Ref.~\cite{Seng:2019lxf}. First, the UV-divergences in the electroweak
sector are reabsorbed into the respective coupling constants and mass parameters following the on-shell
renormalization scheme~\cite{Sirlin:1980nh}. Next, since here we are only interested in the LECs that
do not involve the lepton-hadron interaction, we can take $G_F\to 0$ so the leptons completely decouple
with the quarks. We then retain only the non-leptonic (denoted by the subscript ``nl'') piece in the
Lagrangian that reads:
\begin{equation}
\mathcal{L}_\mathrm{nl}=\mathcal{L}_\mathrm{QCD}-e\bar{\psi}Q_\mathrm{em}\slashed{\mathcal{A}}_<\psi
-\frac{1}{4}F_{\mu\nu}^<F^{\mu\nu}_<-\frac{1}{2\xi}(\partial_\mu \mathcal{A}_<^\mu)^2
+\frac{1}{2}M_\gamma^2\mathcal{A}_\mu^<\mathcal{A}_<^\mu~,
\end{equation}
with $\psi=(u,d,s)^\mathrm{T}$, and $\xi=1$ for the Feynman gauge. Here $\mathcal{A}_<^\mu$ represents the
photon field with its propagator being multiplied by a Pauli-Villars regulator with $\Lambda=M_W$:
\begin{equation}
D_<^{\mu\nu}(z)=\int\frac{d^4q}{(2\pi)^4}e^{-iq\cdot z}\frac{-ig^{\mu\nu}}{q^2-M_\gamma^2}\frac{M_W^2}{M_W^2-q^2}~.
\end{equation}
This extra regulator comes from the splitting of the full photon propagator in the on-shell
renormalization scheme:
\begin{equation}
\frac{1}{q^2}=\frac{1}{q^2-M_W^2}+\frac{1}{q^2}\frac{M_W^2}{M_W^2-q^2}~.
\end{equation}

To make a connection with the chiral Lagrangian in Sec.~\ref{sec:EFT}, we generalize $\mathcal{L}_\mathrm{nl}$
by introducing external sources $\{l_\mu,r_\mu\}$ and spurion fields $\{q_L,q_R\}$:
\begin{eqnarray}
\tilde{\mathcal{L}}_\mathrm{nl}&=&\mathcal{L}_\mathrm{QCD}+\bar{\psi}_L\gamma^\mu\left(l_\mu+q_L
\mathcal{A}_\mu^<\right)\psi_L+\bar{\psi}_R\gamma^\mu\left(r_\mu+q_R\mathcal{A}_\mu^<\right)\psi_R\nonumber\\
&&-\frac{1}{4}F_{\mu\nu}^<F^{\mu\nu}_<-\frac{1}{2\xi}(\partial_\mu \mathcal{A}_<^\mu)^2+\frac{1}{2}M_\gamma^2
\mathcal{A}_\mu^<\mathcal{A}_<^\mu.
\end{eqnarray}
However, unlike Sec.~\ref{sec:EFT}, here we do not identify the external sources and spurions with the
charge matrices and the fermion bilinears, but rather define $l_\mu=v_\mu-a_\mu$, $r_\mu=v_\mu+a_\mu$, $q_L=q_V-q_A$,
$q_R=q_V+q_A$, and decompose them into flavor octet components:
\begin{equation}
v_\mu=v_\mu^a\frac{\lambda^a}{2},\:\:\:a_\mu=a_\mu^a\frac{\lambda^a}{2},\:\:\:q_V=q_V^a\frac{\lambda^a}{2},
\:\:\:q_A=q_A^a\frac{\lambda^a}{2}~,
\end{equation}
where $\{\lambda^a\}$ are the Gell-Mann matrices. We may also define flavor-octet vector and axial currents as:
\begin{equation}
V_\mu^a=\bar{\psi}\gamma_\mu\frac{\lambda^a}{2}\psi,\:\:\:A_\mu^a=\bar{\psi}\gamma_\mu\gamma_5\frac{\lambda^a}{2}\psi~.
\end{equation}
Thus we can write:
\begin{eqnarray}
\tilde{\mathcal{L}}_\mathrm{nl}&=&\mathcal{L}_\mathrm{QCD}+V^{a\mu}\left(v_\mu^a+q_V^a\mathcal{A}_\mu^<\right)
+A^{a\mu}\left(a_\mu^a+q_A^a\mathcal{A}_\mu^<\right)\nonumber\\
&&-\frac{1}{4}F_{\mu\nu}^<F^{\mu\nu}_<-\frac{1}{2\xi}(\partial_\mu \mathcal{A}_<^\mu)^2+\frac{1}{2}M_\gamma^2
\mathcal{A}_\mu^<\mathcal{A}_<^\mu~.\label{eq:Ltildenl}
\end{eqnarray}

\subsection{Defining the four-point correlation functions}

Using the generating functional of the action $\tilde{S}_\mathrm{nl}$ in the presence of external sources
and spurions:
\begin{equation}
W(v,a,q_V,q_A)=\frac{1}{Z}\int D(\bar{\psi},\psi,\mathcal{A}_<)\exp\{i\tilde{S}_\mathrm{nl}(v,a,q_V,q_A)\},
\end{equation}
we can define three types of four-point correlation functions~\cite{Ananthanarayan:2004qk}:
\begin{eqnarray}
\left\langle A_\alpha^a A_\beta^bQ_V^cQ_V^d\right\rangle&\equiv&\int d^4xd^4yd^4z e^{ik\cdot y}\left.
\frac{\delta^4 W(v,a,q_V,q_A)}{\delta a^{a\alpha}(x)\delta a^{b\beta}(y)\delta q_V^c(z)\delta q_V^d(0)}\right|_0~,
\nonumber\\
\left\langle A_\alpha^a A_\beta^bQ_A^cQ_A^d\right\rangle&\equiv&\int d^4xd^4yd^4z e^{ik\cdot y}\left.
\frac{\delta^4 W(v,a,q_V,q_A)}{\delta a^{a\alpha}(x)\delta a^{b\beta}(y)\delta q_A^c(z)\delta q_A^d(0)}\right|_0~,
\nonumber\\
\left\langle V_\alpha^a V_\beta^bQ_V^cQ_V^d\right\rangle&\equiv&\int d^4xd^4yd^4z e^{ik\cdot y}\left.
\frac{\delta^4 W(v,a,q_V,q_A)}{\delta v^{a\alpha}(x)\delta v^{b\beta}(y)\delta q_V^c(z)\delta q_V^d(0)}\right|_0~,
\label{eq:4point}
\end{eqnarray}
where $k$ is a freely-chosen external momentum. The ``$|_0$'' means that we take $v_\mu=a_\mu=q_V=q_A=0$ after
the functional derivative, which decouples the quarks from the photon. Obviously, the only possible
Lorentz structures of these correlation functions are $g_{\alpha\beta}$ and $k_\alpha k_\beta$.

Using Eq.~\eqref{eq:Ltildenl}, it is straightforward to show that the correlation functions
above can be written as:
\begin{eqnarray}
    \label{eq:AAQQ}
\left\langle A_\alpha^aA_\beta^bQ_V	^cQ_V^d\right\rangle &=& \int d^4xd^4yd^4z e^{ik\cdot y}\left\langle 0\right|T\{A_\alpha^a(x)A_\beta^b(y)V_\rho^c(z)V_\sigma^d(0)\}\left|0\right\rangle D_<^{\rho\sigma}(z)~,\nonumber\\
\left\langle A_\alpha^aA_\beta^bQ_A	^cQ_A^d\right\rangle &=& \int d^4xd^4yd^4z e^{ik\cdot y}\left\langle 0\right|T\{A_\alpha^a(x)A_\beta^b(y)A_\rho^c(z)A_\sigma^d(0)\}\left|0\right\rangle D_<^{\rho\sigma}(z)~,\nonumber\\
\left\langle V_\alpha^aV_\beta^bQ_V	^cQ_V^d\right\rangle &=& \int d^4xd^4yd^4z e^{ik\cdot y}\left\langle 0\right|T\{V_\alpha^a(x)V_\beta^b(y)V_\rho^c(z)V_\sigma^d(0)\}\left|0\right\rangle D_<^{\rho\sigma}(z)~.\label{eq:QCDME}
\end{eqnarray}
Note that $\left\langle 0\right|T\{...\}\left|0\right\rangle$ are pure QCD matrix elements, so the
RHS of the equations above are in principle calculable on the lattice.  For instance, the
hadronic part in the correlation functions defined in Eq.~(\ref{eq:AAQQ}) can be
calculated using the sequential-source propagators. Combining the
hadronic part with the photonic weight function of $D_<^{\rho\sigma}(z)$, the whole
4-point correlation functions can be constructed in  lattice simulations.
%There is no need to introduce dynamical photons on the lattice because the photonic effects are all contained in $D_<^{\rho\sigma}(z)$.

\begin{figure}
	\begin{centering}
		\includegraphics[scale=0.1]{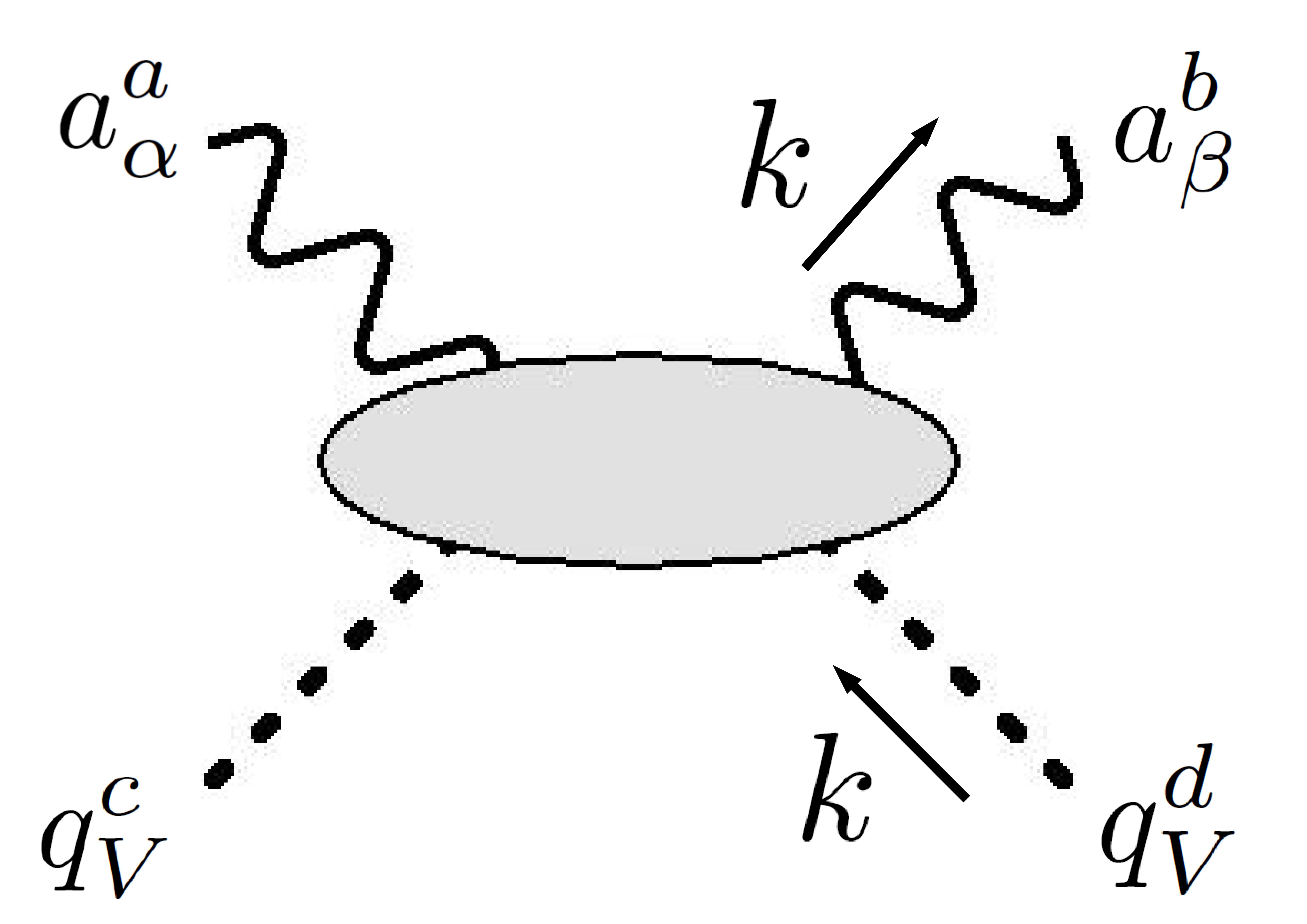}\hfill
		\par\end{centering}
	        \caption{\label{fig:diagrammatic}Diagrammatic representation of $\left\langle A_\alpha^aA_\beta^b Q_V^c
                  Q_V^d\right\rangle$. The other correlation functions can be represented in a similar way.}
\end{figure}

There is a simple diagrammatic interpretation of the correlation functions. Take $\left\langle A_\alpha^a A_\beta^b
Q_V^c Q_V^d\right\rangle$ as an example: It is nothing but the amplitude $iM(q_V^c(0)q_V^d(k)\to a_\alpha^a(0)
a_\beta^b(k))$ calculated using the action $\tilde{S}_\mathrm{nl}(v,a,q_V,q_A)$, see Fig.~\ref{fig:diagrammatic}
(notice that $v,a,q_V,q_A$ are not dynamical fields and do not propagate internally). Therefore, the strategy
is to make use of the ChPT representation of $\tilde{S}_\mathrm{nl}$ to calculate the correlation functions. 
The results obviously depend on the unknown LECs $\{K_i^r\}$. Comparing the ChPT expression and the lattice
calculation of the correlation functions then allows us to determine the unknown LECs. 

Before proceeding with the ChPT calculation, we make a final comment on the correlation functions in
Eq.~\eqref{eq:QCDME}. Due to the existence of the Pauli-Villars regulator in $D^{\rho\sigma}_<(z)$, all the
space-time integrals with respect to $x,y,z$ are convergent. Still, if the LECs probe the
physics at the scale $q\sim M_W$, then the corresponding correlation functions are not fully computed by
lattice QCD alone because this will require a lattice spacing of the size $a\sim 1/M_W$ which is not
achievable in practice. Fortunately, unlike $K_{12}^r$ (see the discussion in Sec.~\ref{sec:EFT}), none of the LECs $K_1^r,...,K_6^r$ is sensitive to physics at
the UV-scale, so the use of a typical lattice spacing is sufficient.

\subsection{ChPT representation of the four-point functions}

The four-point functions defined in Eq.~\eqref{eq:4point} were already calculated in ChPT to $\mathcal{O}(e^2p^2)$
in Ref.~\cite{Ananthanarayan:2004qk}, but there they worked in the chiral limit and retained only
the $g_{\alpha\beta}$ structure, making the results not directly applicable for the lattice. Here, we redo
the calculation at the SU(3) symmetric point with non-zero $M_\phi$ and include both the $g_{\alpha\beta}$ and
$k_\alpha k_\beta$ structures. Following that reference, we cast our results in terms of the four flavor
basis defined below:
\begin{eqnarray}
\hat{e}_1&=&f^{acg}f^{bdg}+f^{adg}f^{bcg}~,\nonumber\\
\hat{e}_2&=&\delta^{ac}\delta^{bd}+\delta^{ad}\delta^{bc}~,\nonumber\\
\hat{e}_3&=&d^{acg}d^{bdg}+d^{adg}d^{bcg}~,\nonumber\\
\hat{e}_4&=&f^{abg}f^{cdg}.
\end{eqnarray}

\begin{figure}
	\begin{centering}
		\includegraphics[scale=0.08]{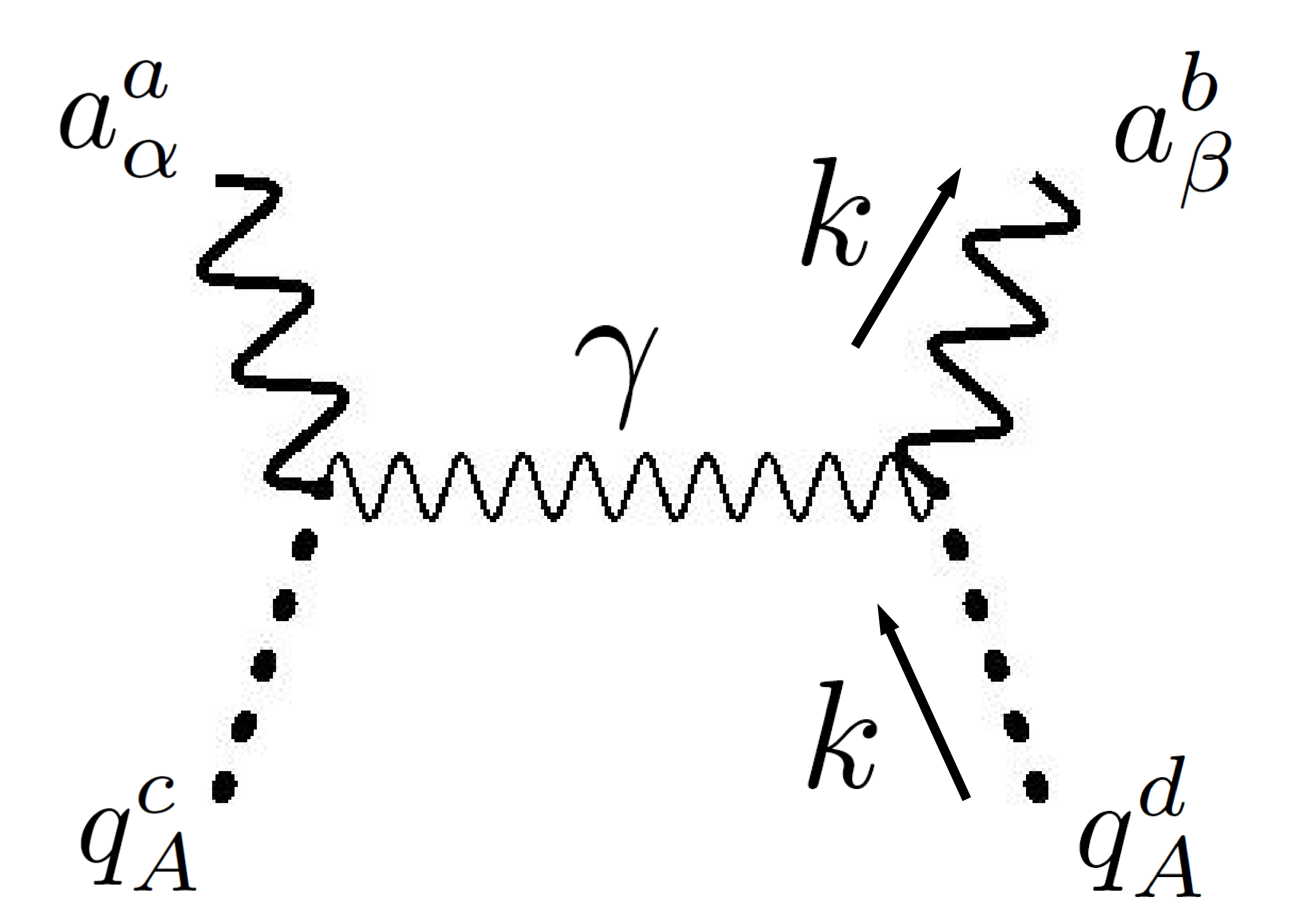}
		\includegraphics[scale=0.08]{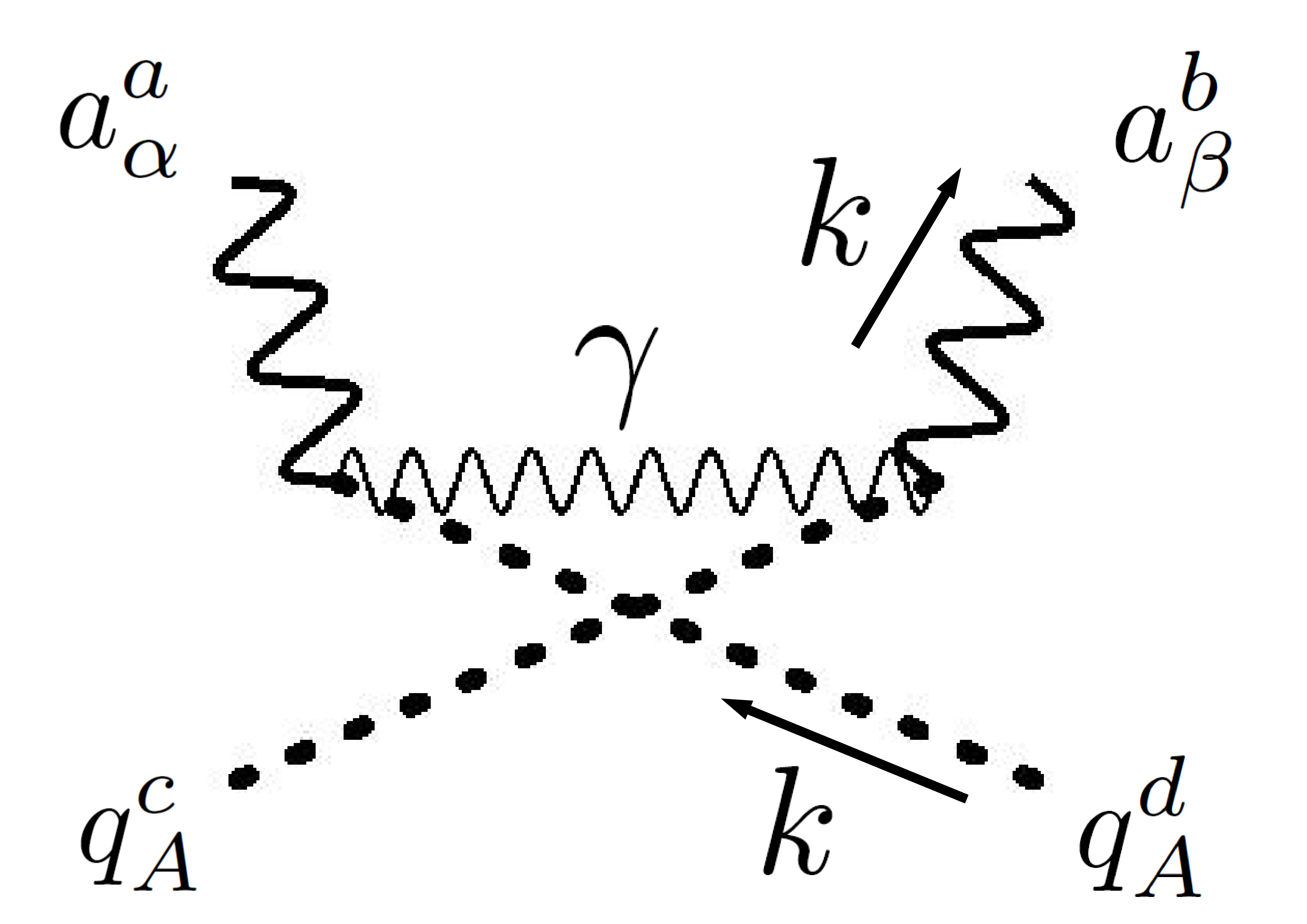}\hfill
		\par\end{centering}
	        \caption{\label{fig:Op2}$\mathcal{O}(p^2)$ contributions to $\left\langle A_\alpha^aA_\beta^b
                     Q_A^cQ_A^d\right\rangle$.}
\end{figure}

\begin{figure}
	\begin{centering}
		\includegraphics[scale=0.08]{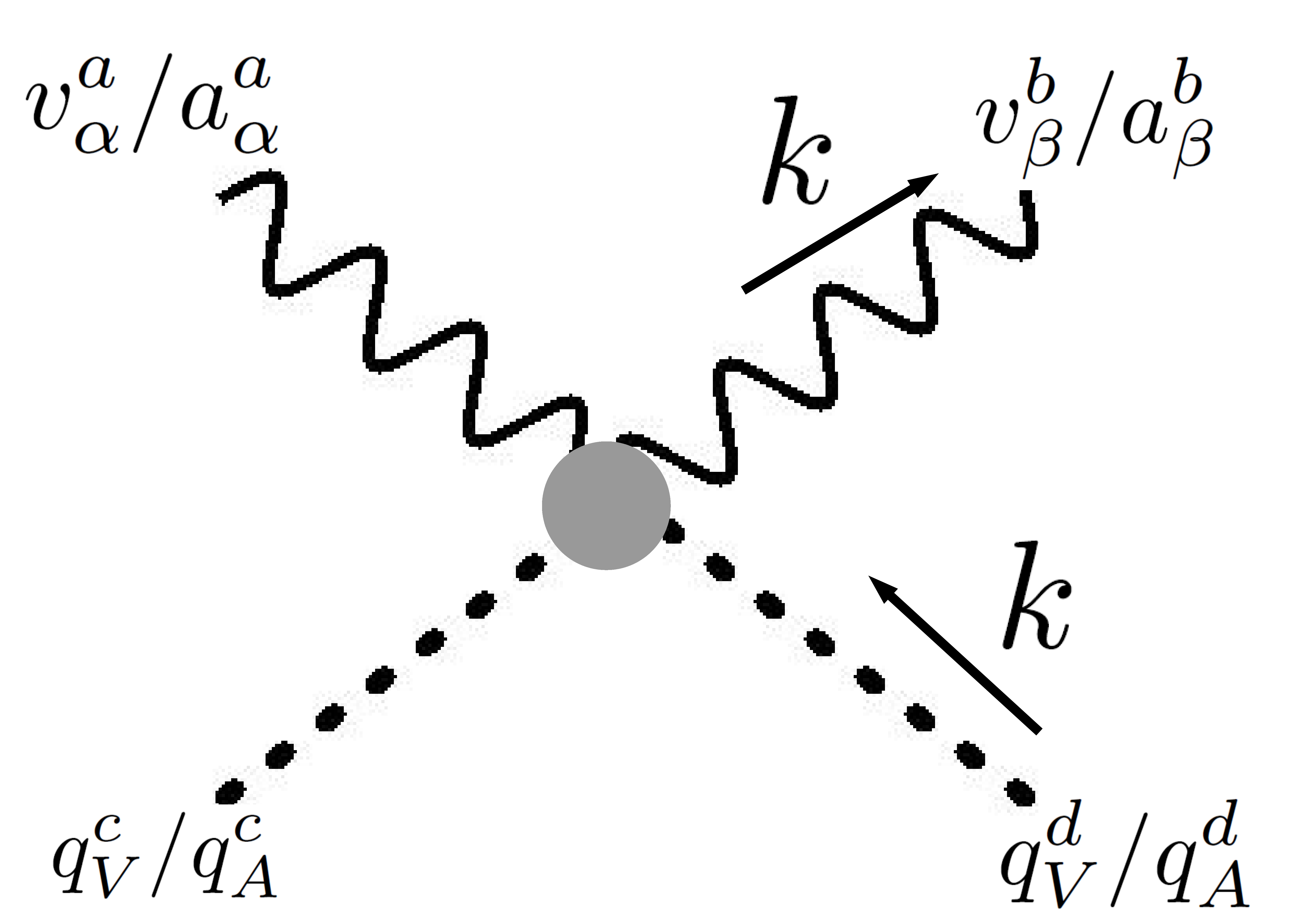}
		\includegraphics[scale=0.08]{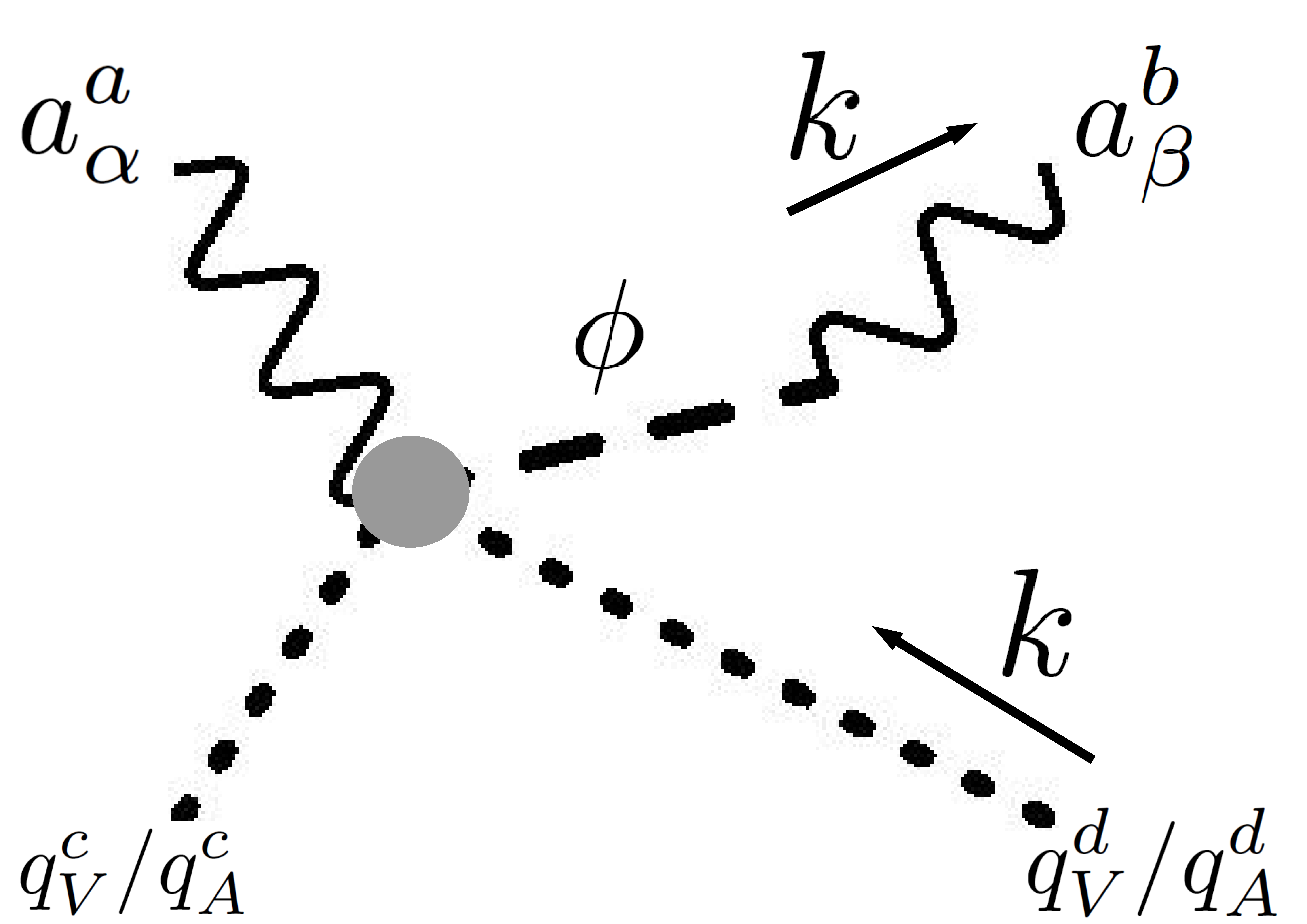}\hfill
		\par\end{centering}
	        \caption{\label{fig:Ki} LEC contributions to the correlation functions. The gray dot represents
                  the counterterm vertex.}
\end{figure}

Up to $\mathcal{O}(e^2p^2)$, the four-point functions read:
\begin{eqnarray}
\left\langle A_\alpha^a A_\beta^b Q_V^c Q_V^d\right\rangle &=& iF_0^2g_{\alpha\beta}\sum_{i=1}^3\alpha_{AV}^{(i)}\hat{e}_i+iF_0^2\frac{k_\alpha k_\beta}{k^2-M_\phi^2}\sum_{i=1}^4\beta_{AV}^{(i)}\hat{e}_i\nonumber\\
&&+\left\langle A_\alpha^a A_\beta^b Q_V^c Q_V^d\right\rangle_\phi+\left\langle A_\alpha^a A_\beta^b Q_V^c Q_V^d\right\rangle_\gamma~,\nonumber\\
\left\langle A_\alpha^a A_\beta^b Q_A^c Q_A^d\right\rangle &=& iF_0^2g_{\alpha\beta}\left[\delta^{ad}\delta^{bc}\frac{F_0^2}{k^2-M_\gamma^2}-\delta^{ac}\delta^{bd}\frac{F_0^2}{M_\gamma^2}\right]+iF_0^2g_{\alpha\beta}\sum_{i=1}^3\alpha_{AA}^{(i)}\hat{e}_i\nonumber\\
&&+iF_0^2\frac{k_\alpha k_\beta}{k^2-M_\phi^2}\sum_{i=1}^4\beta_{AA}^{(i)}\hat{e}_i+\left\langle A_\alpha^a A_\beta^b Q_A^c Q_A^d\right\rangle_\phi+\left\langle A_\alpha^a A_\beta^b Q_A^c Q_A^d\right\rangle_\gamma~,\nonumber\\
\left\langle V_\alpha^a V_\beta^b Q_V^c Q_V^d\right\rangle &=&iF_0^2g_{\alpha\beta}\alpha_{VV}^{(1)}\hat{e}_1+\left\langle V_\alpha^a V_\beta^b Q_V^c Q_V^d\right\rangle_\phi~.\label{eq:EFTresult}
\end{eqnarray}
Let us explain the results above. First, the square bracket in $\left\langle A_\alpha^aA_\beta^bQ_A^cQ_A^d\right
\rangle$ represents the only $\mathcal{O}(p^2)$ contribution that comes from the diagrams shown in
Fig.~\ref{fig:Op2} which is, for some reason, missing in Ref.~\cite{Ananthanarayan:2004qk}. All the
others are $\mathcal{O}(e^2p^2)$. The coefficients $\alpha^{(i)}$ and $\beta^{(i)}$ contain the contributions
from the LECs (as depicted in Fig.~\ref{fig:Ki}) as well as the UV-divergent part of the loop contributions
\footnote{We find that Eq.~(2.15) in Ref.~\cite{Ananthanarayan:2004qk} is wrong by a sign.}. The remaining
parts that carry the subscript $\phi$ and $\gamma$ denote the UV-finite contributions of the meson
and photon loop diagrams, further detail can be found in Appendix~\ref{sec:loops}.

Let us concentrate on the coefficients $\alpha^{(i)}$ and $\beta^{(i)}$. They read:
\begin{eqnarray}
\alpha_{AV}^{(1)}&=&2K_1^r+2K_2^r+2K_5^r+2K_6^r+4K_{12}^r-K_{13}+2K_{14}+\frac{5Z-2}{32\pi^2}\ln\frac{\mu^2}{M_\phi^2}~,
\nonumber\\
\alpha_{AV}^{(2)}&=&-4K_3^r+2K_4^r+\frac{4}{3}K_5^r+\frac{4}{3}K_6^r+\frac{3Z}{16\pi^2}\ln\frac{\mu^2}{M_\phi^2}~,
\nonumber\\
\alpha_{AV}^{(3)}&=&6K_1^r+6K_2^r+2K_5^r+2K_6^r+\frac{9Z}{32\pi^2}\ln\frac{\mu^2}{M_\phi^2}~,
\end{eqnarray}
\begin{eqnarray}
\beta_{AV}^{(1)}&=&-2K_1^r-2K_2^r-2K_5^r-2K_6^r-2K_{12}^r+K_{13}+\frac{5-10Z}{64\pi^2}\ln\frac{\mu^2}{M_\phi^2}~,
\nonumber\\
\beta_{AV}^{(2)}&=&4K_3^r-2K_4^r-\frac{4}{3}K_5^r-\frac{4}{3}K_6^r-\frac{3Z}{16\pi^2}\ln\frac{\mu^2}{M_\phi^2}~,
\nonumber\\
\beta_{AV}^{(3)}&=&-6K_1^r-6K_2^r-2K_5^r-2K_6^r-\frac{9Z}{32\pi^2}\ln\frac{\mu^2}{M_\phi^2}~,\nonumber\\
\beta_{AV}^{(4)}&=&-2K_{12}^r+K_{13}-\frac{1}{64\pi^2}\ln\frac{\mu^2}{M_\phi^2}~,
\end{eqnarray}
\begin{eqnarray}
\alpha_{AA}^{(1)}&=&2K_1^r-2K_2^r+2K_5^r-2K_6^r+4K_{12}^r+K_{13}+2K_{14}-\frac{5Z+2}{32\pi^2}\ln\frac{\mu^2}{M_\phi^2}~,
\nonumber\\
\alpha_{AA}^{(2)}&=&-4K_3^r-2K_4^r+\frac{4}{3}K_5^r-\frac{4}{3}K_6^r-\frac{3Z}{16\pi^2}\ln\frac{\mu^2}{M_\phi^2}~,
\nonumber\\
\alpha_{AA}^{(3)}&=&6K_1^r-6K_2^r+2K_5^r-2K_6^r-\frac{9Z}{32\pi^2}\ln\frac{\mu^2}{M_\phi^2}~,
\end{eqnarray}
\begin{eqnarray}
\beta_{AA}^{(1)}&=&-2K_1^r+2K_2^r-2K_5^r+2K_6^r-2K_{12}^r-K_{13}+\frac{5+10Z}{64\pi^2}\ln\frac{\mu^2}{M_\phi^2}~,\nonumber\\
\beta_{AA}^{(2)}&=&4K_3^r+2K_4^r-\frac{4}{3}K_5^r+\frac{4}{3}K_6^r+\frac{3Z}{16\pi^2}\ln\frac{\mu^2}{M_\phi^2}~,\nonumber\\
\beta_{AA}^{(3)}&=&-6K_1^r+6K_2^r-2K_5^r+2K_6^r+\frac{9Z}{32\pi^2}\ln\frac{\mu^2}{M_\phi^2}~,\nonumber\\
\beta_{AA}^{(4)}&=&-2K_{12}^r-K_{13}-\frac{1}{64\pi^2}\ln\frac{\mu^2}{M_\phi^2}~,
\end{eqnarray}
and finally, $\alpha_{VV}^{(1)}=K_{13}+2K_{14}$. They provide an over-complete set of equations to solve for
the needed LECs, an example of solutions is given in Appendix~\ref{sec:allKi}. So in principle one
could calculate each correlation function with several flavor combinations to extract the needed
coefficients $\alpha^{(i)}$ and $\beta^{(i)}$, and with them one could determine all the $\{K_i^r\}$ individually. 
However, if we are only interested in the unsuppressed combination of $\{K_i^r\}$ that enters
$\delta_\mathrm{em}^{K^\pm l}$ (see Eq.~\eqref{eq:deltaemKl3}), things are much simpler: It can be obtained
from a single four-point function at zero external momentum:
\begin{equation}
\left\langle A_\alpha^1A_\beta^8Q_V^8Q_V^1\right\rangle_{k=0}=iF_0^2g_{\alpha\beta}\left[-4K_3^r+2K_4^r+\frac{4}{3}K_5^r+\frac{4}{3}K_6^r+\frac{3Z}{16\pi^2}\left(-1+\ln\frac{\mu^2}{M_\phi^2}\right)\right]~,\label{eq:4pointmatching}
\end{equation}
which is the last central result of this paper.

This completes the setup of the problem for the future lattice calculation. 
The chiral LEC's are unambiguously related to a 4-point correlation function and the axial $\gamma W$ box.  Using lattice QCD simulations, one can expect to determine the LECs with
controlled uncertainties and provide useful information for the electromagnetic
corrections to $K_{l3}$ decays.

\section{\label{sec:conclusion}Conclusions}

We have entered a new era where lattice QCD becomes increasingly important in the studies of high-precision
electromagnetic effects in low-energy phenomena. In particular, it is now timely to extend its impact to
the field of semileptonic beta decays which plays a decisive role in the precision test of the top-row CKM
matrix unitarity and the implications for BSM physics therein. 

It is expected to be extremely challenging to perform a full lattice QCD calculation to the virtual + real QED corrections to the kaon semileptonic decay rate, of which the estimated time span is of the order of $10^1$ years. Given the current status of the CKM unitarity, it is highly desirable to look for an alternative starting point such that lattice QCD can make immediate impact to the field. In this paper we propose a strategy of such kind. We first point out that, at $\mathcal{O}(e^2p^2)$ in chiral power counting, 
there are only three combinations of LECs that are relevant for $K_{l3}$ and $\pi_{e3}$ decays: $X_1$, $\bar{X}_6^\mathrm{phys}$ and $-2K_3^r+K_4^r+(2/3)(K_5^r+K_6^r)$. Based on a careful comparison between the Sirlin's representation and the ChPT representation of the QED effects,
we show that these LECs can all be pinned down by calculating a few simple quantities on the lattice.

To obtain the LECs $X_1$ and $\bar{X}_6^\mathrm{phys}$, we need to calculate the axial $\gamma W$-box diagrams for the $\pi^0\pi^+$ and $\pi^-K^0$ systems in the degenerate limit. The former was already performed in Ref.~\cite{Feng:2020zdc}, which translates into a determination of $(4/3)X_1+\bar{X}_6^\mathrm{phys}$ with 10\% accuracy. We observe that the outcome is significantly different from the resonant model calculation widely adopted in the existing $K_{l3}$ RC analysis, which adds to the urgency of our proposed calculations. The $\pi^-K^0$ axial box can be computed in exactly the same way, and in fact its result will be available in the near future.

On the other hand, the extraction of the LECs $\{K_i^r\}$ will be based on the lattice calculation of the four-point correlation functions defined in Eq.\eqref{eq:QCDME} which can be done using, e.g., sequential-source propagators. In particular, we show an example in Appendix \ref{sec:allKi} where all individual $\{K_i^r\}$ are obtained from the coefficients $\{\beta^{(i)}\}$ in the four-point functions. In practice it is of course not so trivial, because these coefficients are associated to the $k_\alpha k_\beta$ structure that is sensitive to the direction of the external momentum $k$, which may lead to extra systematic uncertainties due to the breaking of the exact rotational symmetry on the lattice (it is not possible to solve for all individual $\{K_i^r\}$ using only the simpler coefficients $\{\alpha^{(i)}\}$ without imposing further assumptions, such as large-$N_c$ approximation, which one normally avoids in first-principles calculations). Fortunately, as far as the relevant linear combination $-2K_3^r+K_4^r+(2/3)(K_5^r+K_6^r)$ is concerned, one needs only to calculate a single four-point correlation function with zero external momentum, as indicated in Eq.\eqref{eq:4pointmatching}. We will defer the discussions of the actual lattice QCD setup needed for such a calculation to a future work.

Our proposed calculation will not only improve
the precision of the $|V_{us}|$ extraction from $K_{l3}$ alone, but will also reduce the theoretical
uncertainty in the ratio $R_V=\Gamma_{K_{l3}}/\Gamma_{\pi_{e3}}$ that helps us to better understand the
disagreement between the $K_{l2}$ and $K_{l3}$ extractions of $|V_{us}|$.

\section*{Acknowledgements} 

We thank Vincenzo Cirigliano and Bachir Moussallam for many inspiring discussions. This work is supported in
part by  the DFG (Grant No. TRR110)
and the NSFC (Grant No. 11621131001) through the funds provided
to the Sino-German CRC 110 ``Symmetries and the Emergence of
Structure in QCD" (U-G.M and C.Y.S), by the Alexander von Humboldt Foundation through the Humboldt
Research Fellowship (C.Y.S), by the Chinese Academy of Sciences (CAS) through a President's
International Fellowship Initiative (PIFI) (Grant No. 2018DM0034) and by the VolkswagenStiftung
(Grant No. 93562) (U-G.M), by EU Horizon 2020 research and innovation programme, STRONG-2020 project
under grant agreement No 824093 and by the German-Mexican research collaboration Grant No. 278017 (CONACyT)
and No. SP 778/4-1 (DFG) (M.G), by NSFC of China under Grant No. 11775002 (X.F) and by
DOE grant DE-SC0010339 (L.C.J).

\begin{appendix}
	
\section{\label{sec:loops}loop contributions to the four-point functions}	

In this Appendix we present the UV-finite parts of the one-loop contributions to the four-point
correlation functions in Eq.~\eqref{eq:EFTresult}\footnote{We acknowledge the power of Package-X
that provides the fully analytic expressions of all loop integrals in terms of elementary
functions~\cite{Patel:2015tea,Patel:2016fam}.}.

\subsection{$\mathcal{O}(e^2p^2)$ contributions from meson loops}

\begin{figure}
	\begin{centering}
		\includegraphics[scale=0.08]{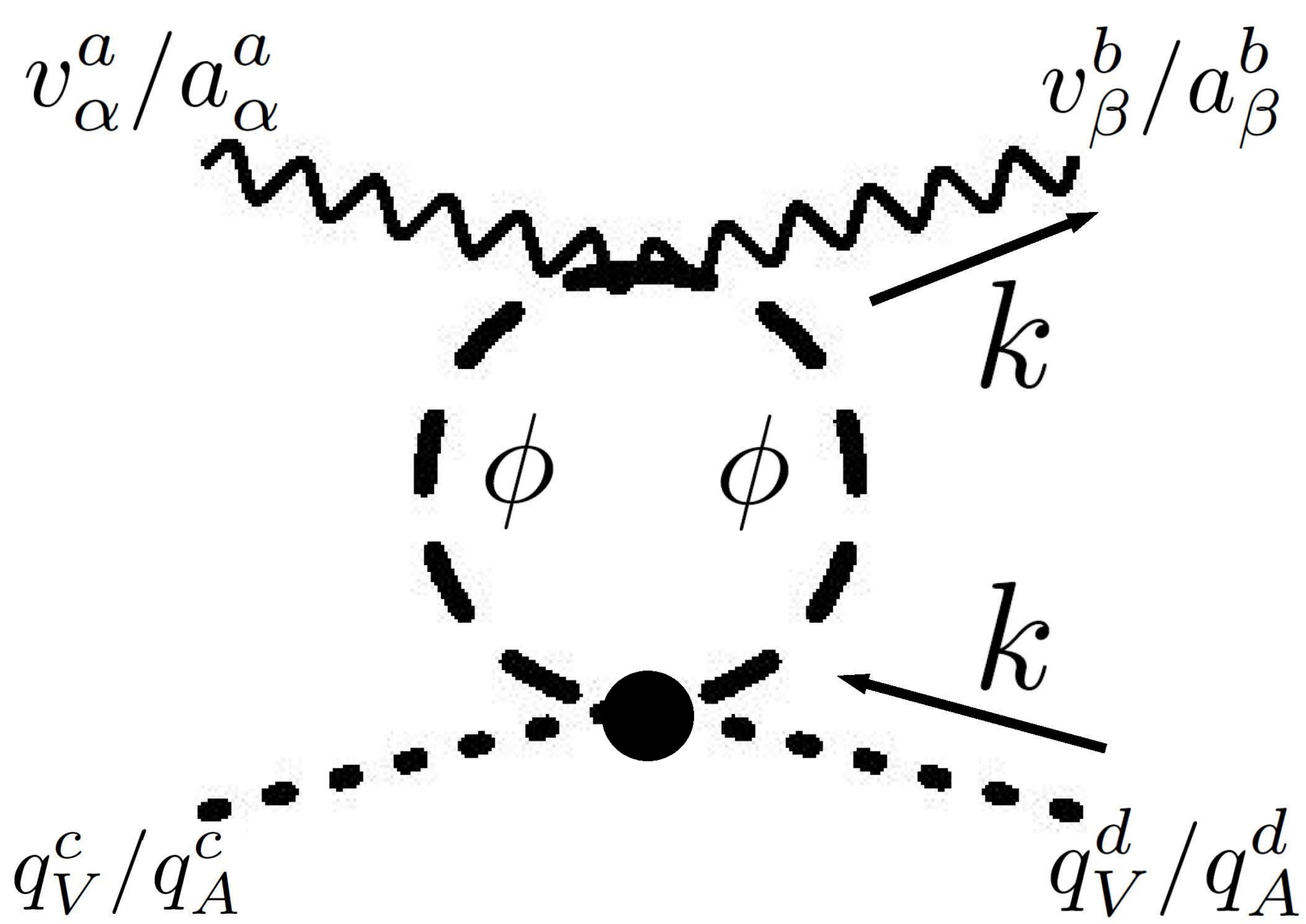}
		\includegraphics[scale=0.08]{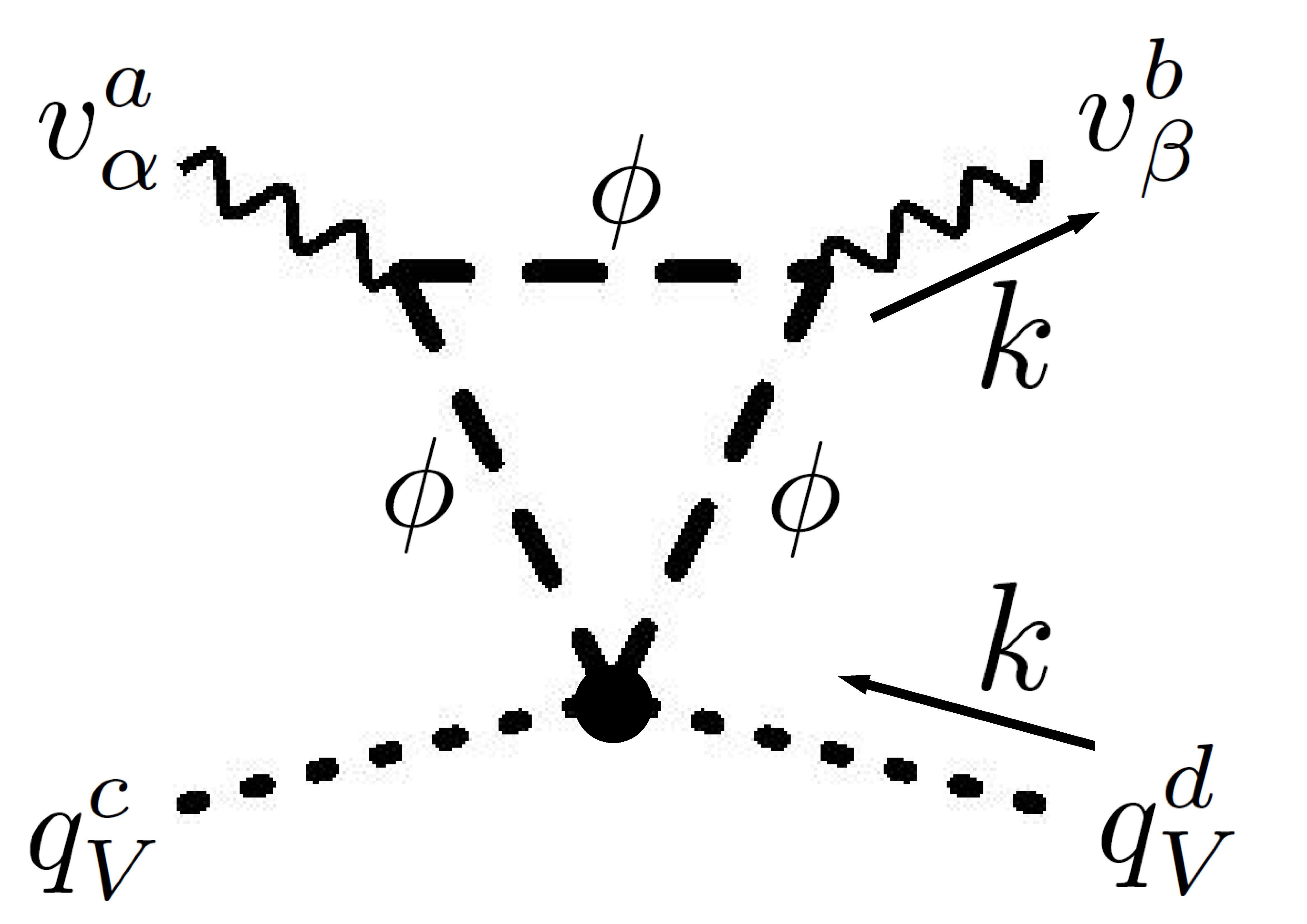}
		\includegraphics[scale=0.08]{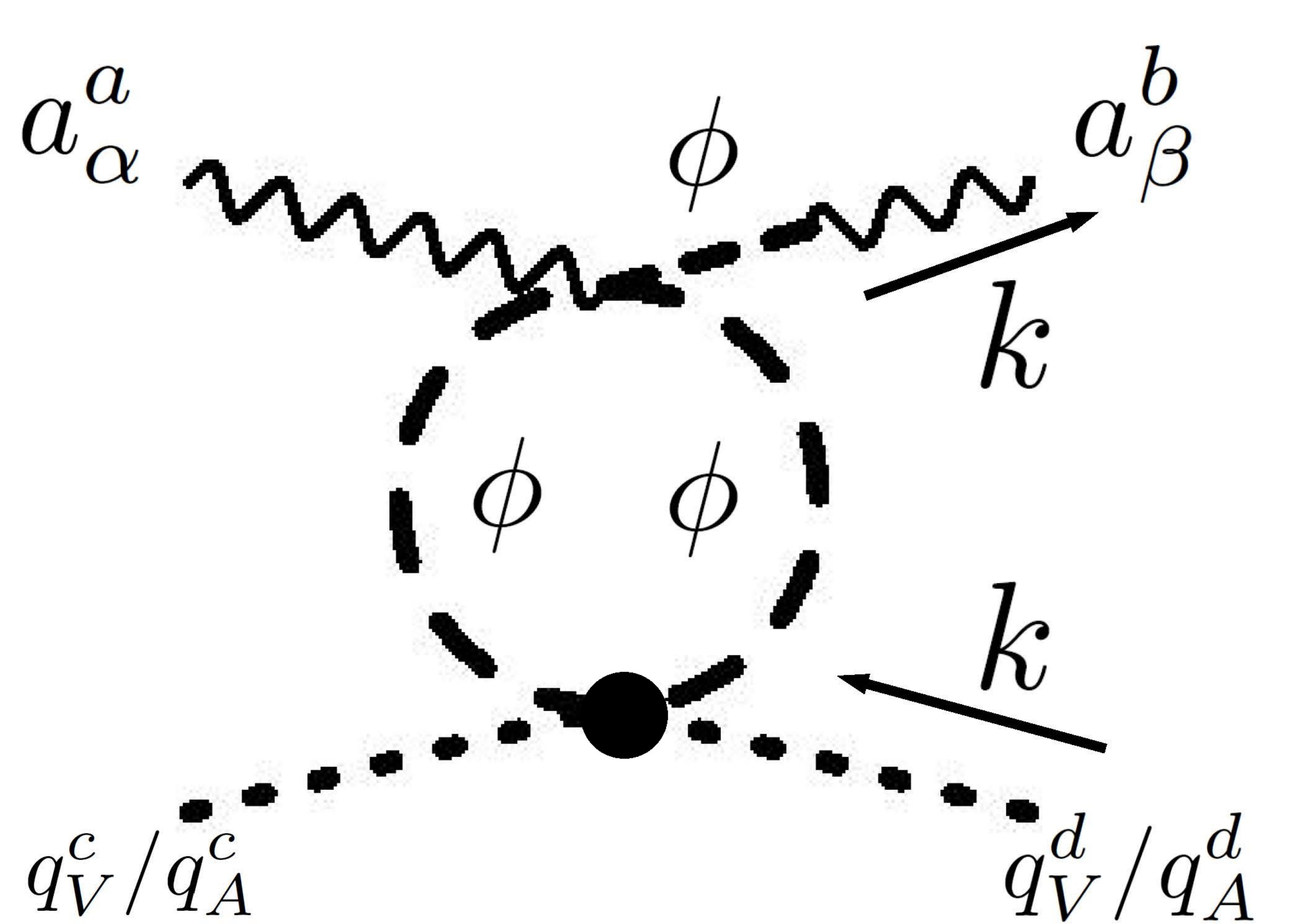}\hfill
		\par\end{centering}
	        \caption{\label{fig:mesonloop}Contributions from meson loops. The black circle denotes the
                  $\mathcal{O}(e^2)$ vertex. The third diagram contains a meson pole.}
\end{figure}

The meson loop contributions are depicted in Fig.~\ref{fig:mesonloop}. The results are:
\begin{eqnarray}
\left\langle A_\alpha^a A_\beta^bQ_V^cQ_V^d\right\rangle_\phi &=&-\left\langle A_\alpha^a A_\beta^bQ_A^cQ_A^d\right\rangle_\phi\nonumber\\
&=&iF_0^2\frac{Z}{16\pi^2}\left(\frac{5}{2}\hat{e}_1+3\hat{e}_2+\frac{9}{2}\hat{e}_3\right)\left(1+\Lambda(k^2,M_\phi)\right)\left(g_{\alpha\beta}-\frac{k_\alpha k_\beta}{k^2-M_\phi^2}\right)\nonumber\\
\left\langle V_\alpha^aV_\beta^bQ_V^cQ_V^d\right\rangle_\phi &=&0,
\end{eqnarray}
where 
\begin{equation}
\Lambda(k^2,M_\phi)=\frac{\sqrt{k^2(k^2-4M_\phi^2)}}{k^2}\ln\left(\frac{\sqrt{k^2(k^2-4M_\phi^2)}-k^2+2M_\phi^2}{2M_\phi^2}\right).
\end{equation}

\subsection{$\mathcal{O}(e^2p^2)$ contributions from photon loops}

The photon loop contributions involve more Feynman diagrams, so for the benefits of future cross-check,
we split them into two pieces: $\left\langle...\right\rangle_\gamma=\left\langle...\right\rangle_{\gamma_1}
+\left\langle...\right\rangle_{\gamma_2}$, where the two terms on the RHS denote contribution without and with
a meson pole, respectively.

\subsubsection{without meson pole}

\begin{figure}
	\begin{centering}
		\includegraphics[scale=0.08]{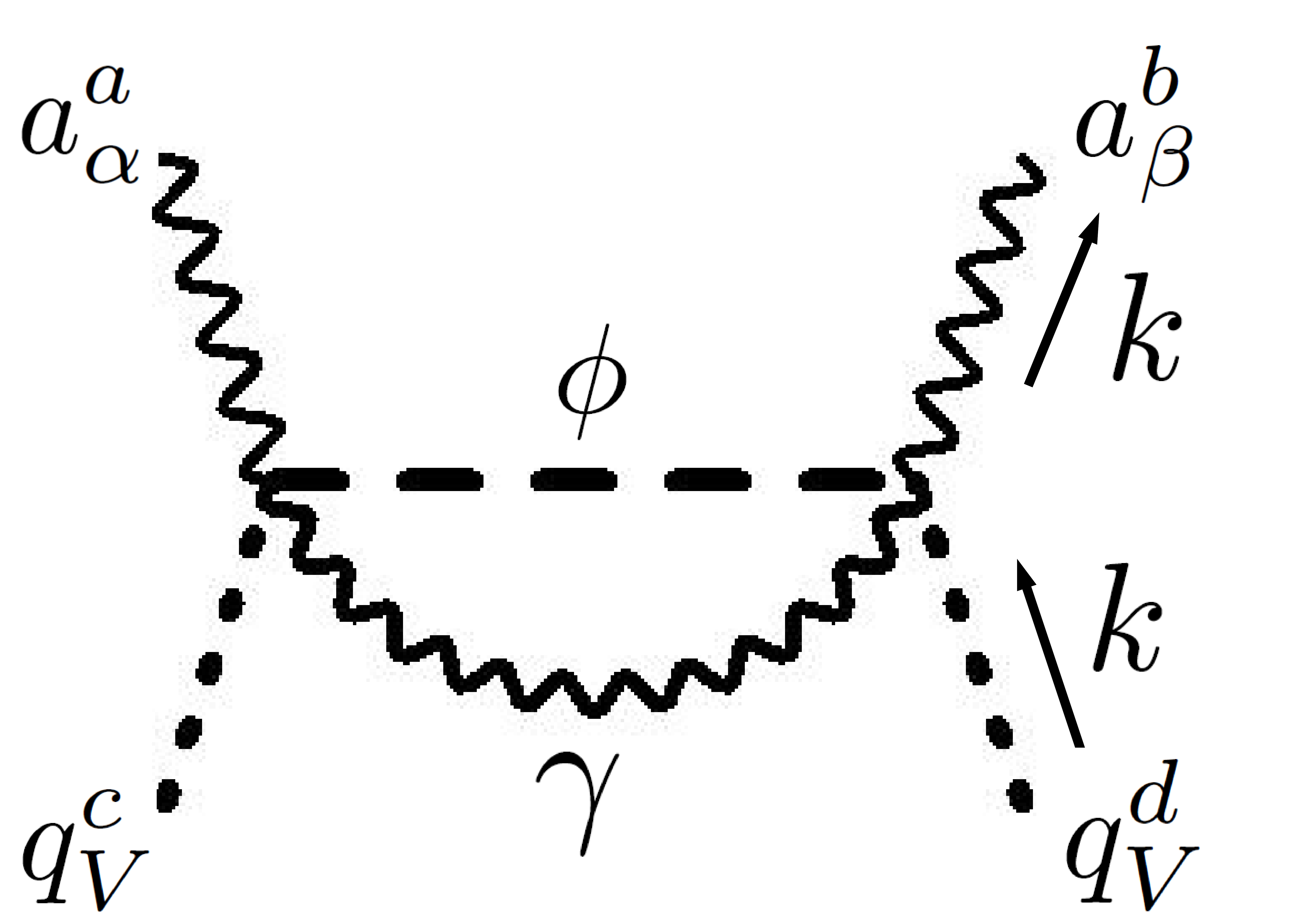}
		\includegraphics[scale=0.08]{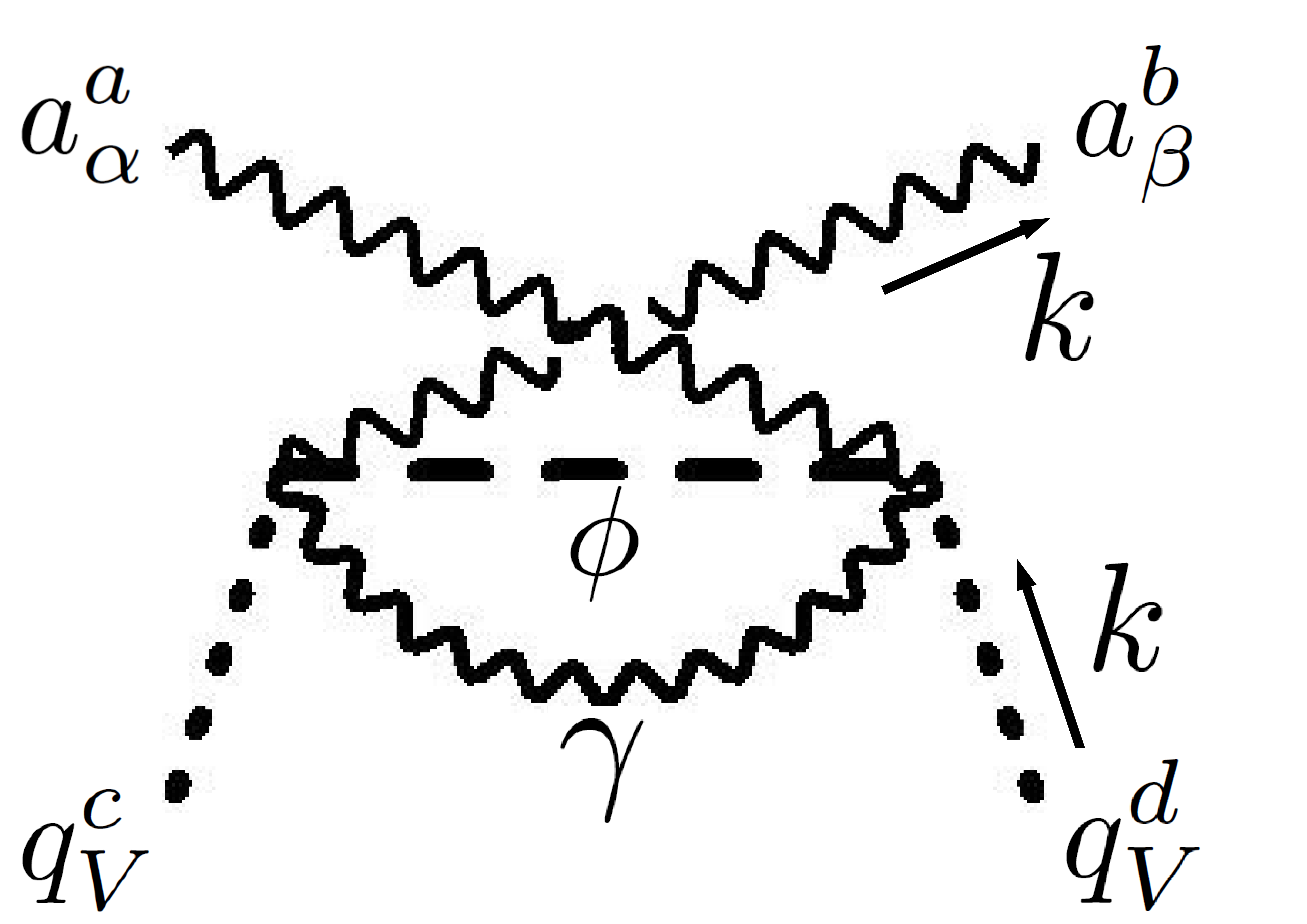}
		\includegraphics[scale=0.08]{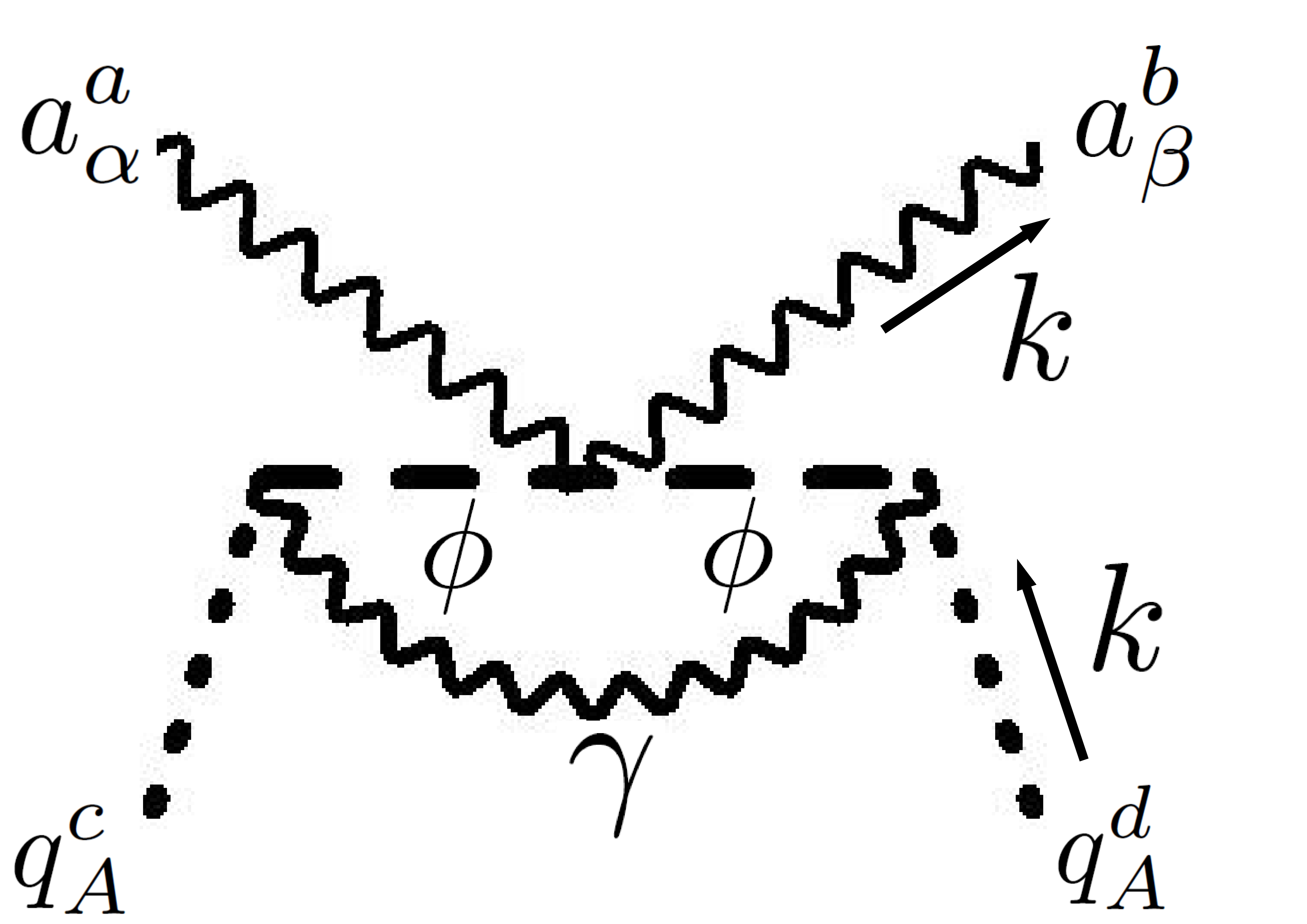}\hfill
		\par\end{centering}
	\caption{\label{fig:photonWOpole}Contributions from photon loops without a meson pole.}
\end{figure}

The photon loop contributions without a meson pole are depicted in Fig.~\ref{fig:photonWOpole}. The results read:
\begin{eqnarray}
\left\langle A_\alpha^a A_\beta^b Q_V^cQ_V^d\right\rangle_{\gamma_1} 
&=&-iF_0^2\frac{1}{16\pi^2}\frac{\hat{e}_1-\hat{e}_4}{2}\left(\frac{k^2-M_\phi^2}{k^2}\ln\frac{M_\phi^2}{M_\phi^2-k^2}+1\right)g_{\alpha\beta}\nonumber\\
\left\langle A_\alpha^a A_\beta^b Q_A^cQ_A^d\right\rangle_{\gamma_1} 
&=&-iF_0^2\frac{1}{16\pi^2}\hat{e}_1\left(\frac{2M_\phi^2-k^2}{2M_\phi^2}\Lambda(k^2,M_\phi)+\frac{(k^2-M_\phi^2)^2}{2k^2M_\phi^2}\ln\frac{M_\phi^2}{M_\phi^2-k^2}+\frac{1}{2}\right)g_{\alpha\beta}.\nonumber\\
\end{eqnarray}

\subsubsection{with meson pole}

\begin{figure}
	\begin{centering}
		\includegraphics[scale=0.08]{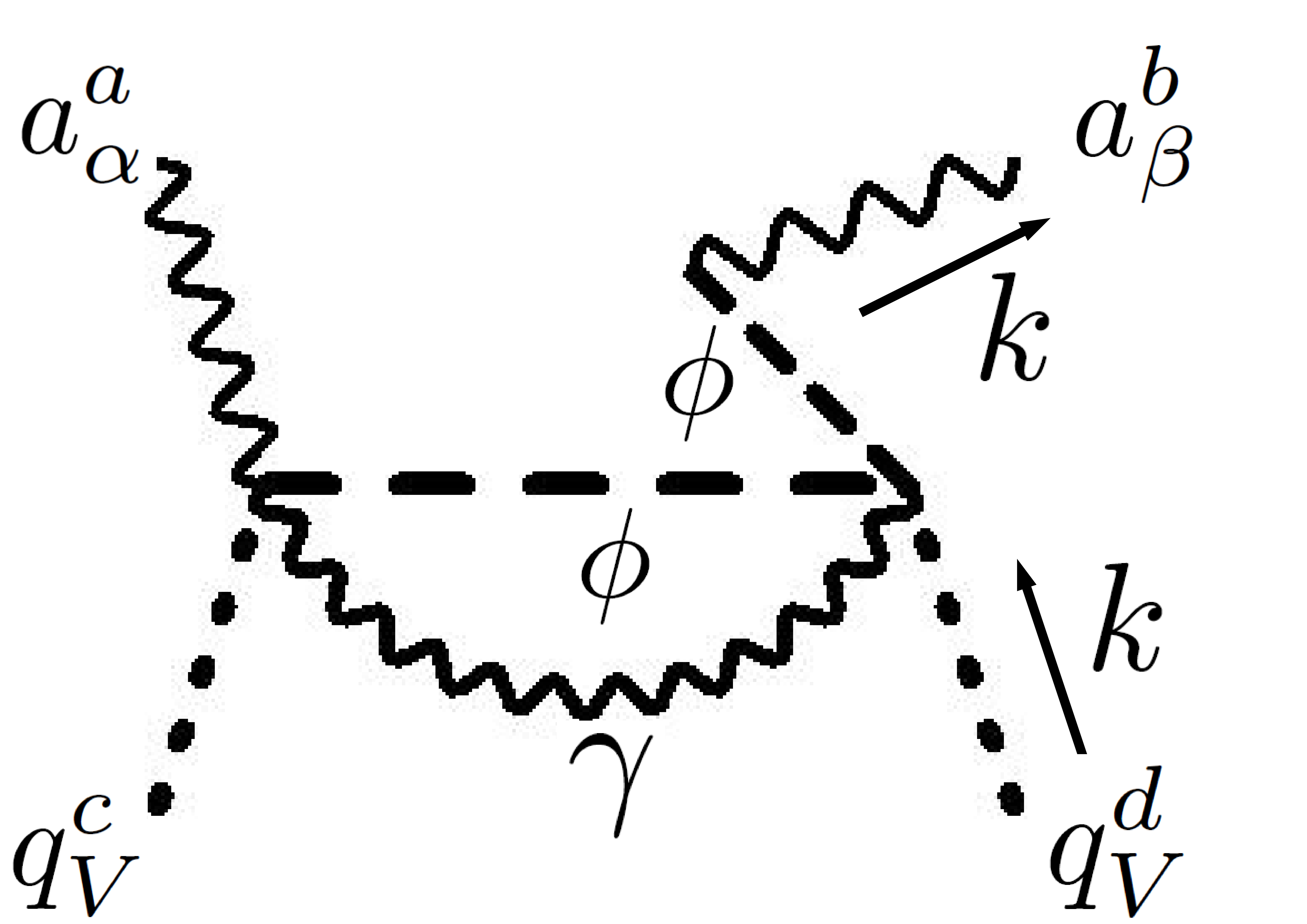}
		\includegraphics[scale=0.08]{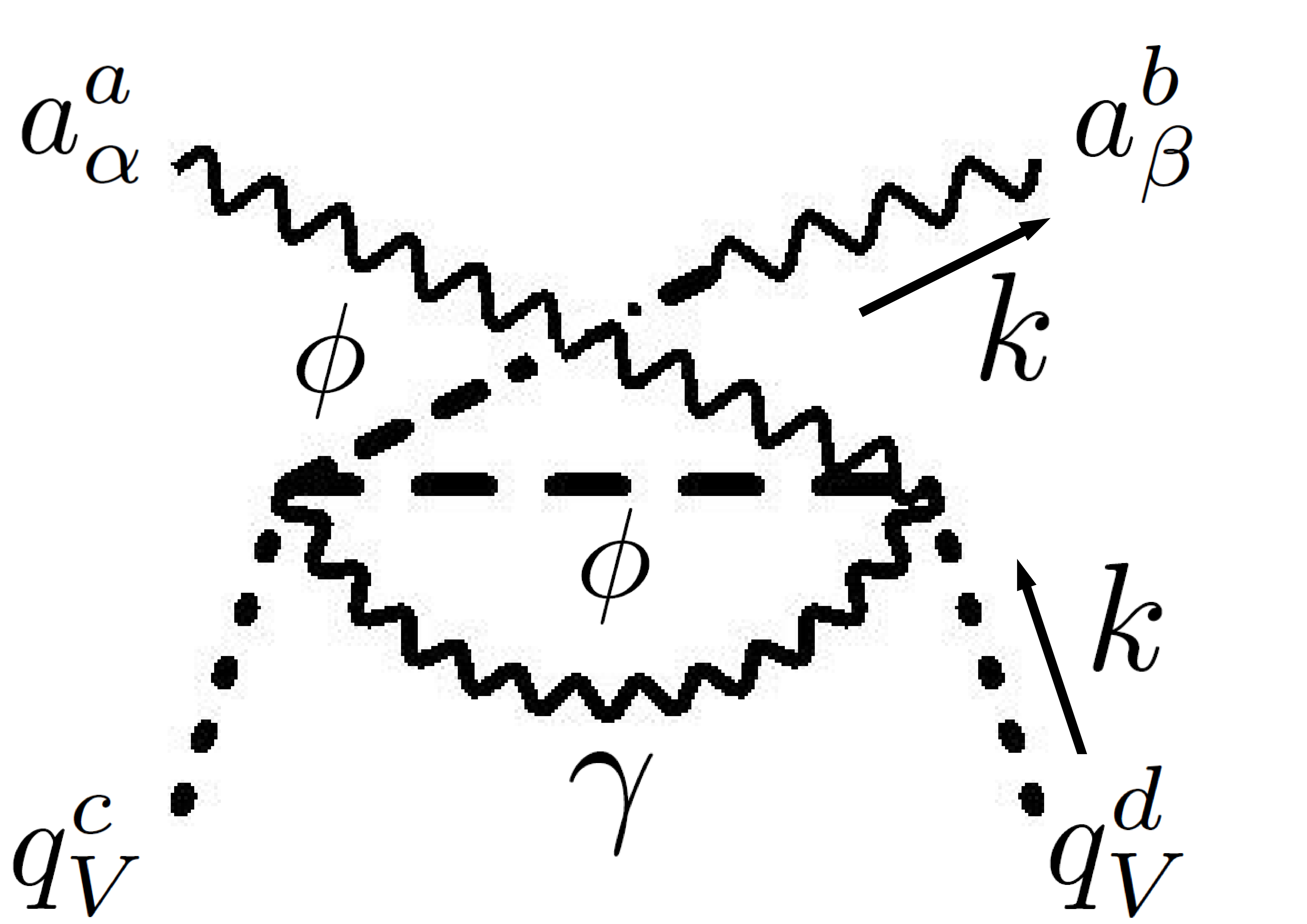}
		\includegraphics[scale=0.08]{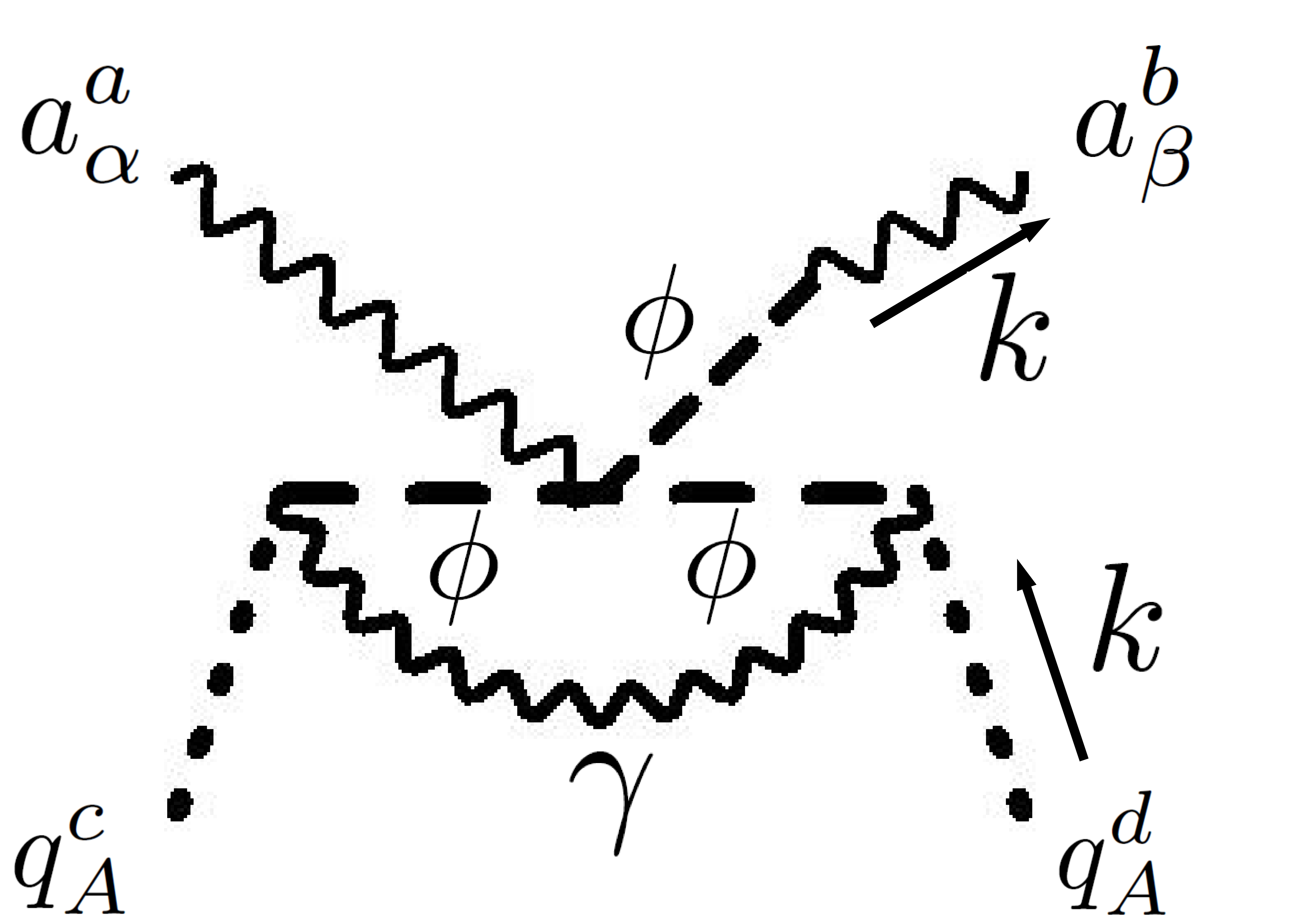}
		\includegraphics[scale=0.08]{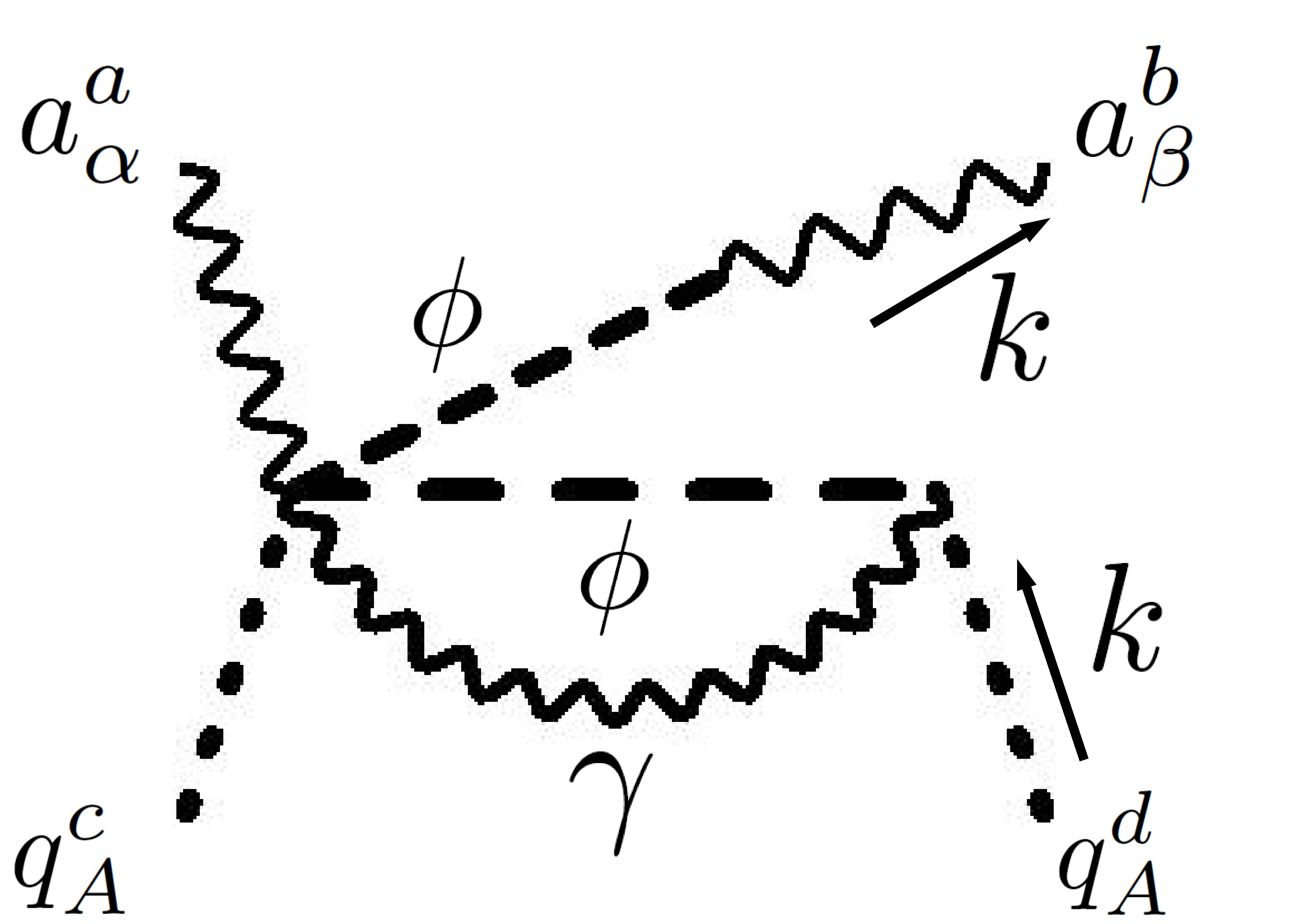}\hfill
		\par\end{centering}
	\caption{\label{fig:photonWpole}Contributions from photon loops with a meson pole.}
\end{figure}

The contributions from photon loops with a meson pole are depicted in Fig.~\ref{fig:photonWpole}. The results read:
\begin{eqnarray}
\left\langle A_\alpha^aA_\beta^bQ_V^cQ_V^d\right\rangle_{\gamma_2} &=&iF_0^2\frac{1}{16\pi^2}\frac{\hat{e}_1-\hat{e}_4}{2}\frac{3k^2+M_\phi^2}{2k^2}\left(\frac{k^2-M_\phi^2}{k^2}\ln\frac{M_\phi^2}{M_\phi^2-k^2}+1\right)\frac{k_\alpha k_\beta}{k^2-M_\phi^2}\nonumber\\
\left\langle A_\alpha^aA_\beta^bQ_A^cQ_A^d\right\rangle_{\gamma_2} &=&iF_0^2\frac{1}{16\pi^2}\left\{\hat{e}_1\left[\frac{2M_\phi^2-k^2}{2M_\phi^2}\Lambda(k^2,M_\phi)+\frac{2(k^2)^3-3(k^2)^2M_\phi^2+2k^2M_\phi^4-M_\phi^6}{4(k^2)^2M_\phi^2}\right.\right.\nonumber\\
&&\left.\times\ln\frac{M_\phi^2}{M_\phi^2-k^2}+\frac{3k^2+M_\phi^2}{4k^2}\right]+\hat{e}_4\left[-\frac{(k^2)^2+4k^2M_\phi^2-5M_\phi^4}{4(k^2)^2}\ln\frac{M_\phi^2}{M_\phi^2-k^2}\right.\nonumber\\
&&\left.\left.-\frac{k^2+5M_\phi^2}{4k^2}\right]\right\}\frac{k_\alpha k_\beta}{k^2-M_b^2}.
\end{eqnarray}

\section{\label{sec:allKi}Obtaining every $K_i^r$ individually}

In this Appendix we present one (out of the many possible) set of solutions for $K_1^r,...,K_6^r$ in terms
of the coefficients $\{\alpha^{(i)},\beta^{(i)}\}$ defined in Eq.~\eqref{eq:EFTresult}. Here we make use
of only $\{\beta^{(i)}\}$:
\begin{eqnarray}
K_1^r&=&\frac{1}{8}\left(\beta_{AA}^{(1)}-\beta_{AA}^{(3)}-\beta_{AA}^{(4)}+\beta_{AV}^{(1)}-\beta_{AV}^{(3)}-\beta_{AV}^{(4)}-\frac{3}{16\pi^2}\ln\frac{\mu^2}{M_\phi^2}\right)\nonumber\\
K_2^r&=&\frac{1}{8}\left(-\beta_{AA}^{(1)}+\beta_{AA}^{(3)}+\beta_{AA}^{(4)}+\beta_{AV}^{(1)}-\beta_{AV}^{(3)}-\beta_{AV}^{(4)}-\frac{Z}{4\pi^2}\ln\frac{\mu^2}{M_\phi^2}\right)\nonumber\\
K_3^r&=&\frac{1}{24}\left(-3\beta_{AA}^{(1)}+3\beta_{AA}^{(2)}+\beta_{AA}^{(3)}+3\beta_{AA}^{(4)}-3\beta_{AV}^{(1)}+3\beta_{AV}^{(2)}+\beta_{AV}^{(3)}+3\beta_{AV}^{(4)}+\frac{9}{16\pi^2}\ln\frac{\mu^2}{M_\phi^2}\right)\nonumber\\
K_4^r&=&\frac{1}{12}\left(-3\beta_{AA}^{(1)}+3\beta_{AA}^{(2)}+\beta_{AA}^{(3)}+3\beta_{AA}^{(4)}+3\beta_{AV}^{(1)}-3\beta_{AV}^{(2)}-\beta_{AV}^{(3)}-3\beta_{AV}^{(4)}-\frac{3Z}{4\pi^2}\ln\frac{\mu^2}{M_\phi^2}\right)\nonumber\\
K_5^r&=&\frac{1}{8}\left(-3\beta_{AA}^{(1)}+\beta_{AA}^{(3)}+3\beta_{AA}^{(4)}-3\beta_{AV}^{(1)}+\beta_{AV}^{(3)}+3\beta_{AV}^{(4)}+\frac{9}{16\pi^2}\ln\frac{\mu^2}{M_\phi^2}\right)\nonumber\\
K_6^r&=&\frac{1}{8}\left(3\beta_{AA}^{(1)}-\beta_{AA}^{(3)}-3\beta_{AA}^{(4)}-3\beta_{AV}^{(1)}+\beta_{AV}^{(3)}+3\beta_{AV}^{(4)}-\frac{3Z}{8\pi^2}\ln\frac{\mu^2}{M_\phi^2}\right).\label{eq:allLECs}
\end{eqnarray}

\end{appendix}

%\bibliographystyle{prsty}
%\bibliography{formfactor_ref}

%%%%%%%%%%%%%%%%%%%%%%%%%%%%%%%%%%%%%%%%%%%%%%%%%%%%%%%%%%
%\begin{thebibliography}{11} 

\providecommand{\href}[2]{#2}\begingroup\raggedright\endgroup

\end{document}